\documentclass[12pt]{amsart}

\newtheorem{thm}{Theorem}[section]

\newtheorem{cor}[thm]{Corollary}
\newtheorem{lem}[thm]{Lemma}
\newtheorem{prop}[thm]{Proposition}
\newtheorem*{prob*}{Problem}

\newtheorem*{thm*}{Theorem}

\theoremstyle{definition}
\newtheorem{defn}[thm]{Definition}

\newtheorem*{defn*}{Definition}
\newtheorem{rem}[thm]{Remark}
\numberwithin{equation}{section}

\newcommand{\Epsilon}{\mathcal{E}}
\newcommand{\Dc}{\mathcal{D}}
\newcommand{\C}{\mathbb C}
\newcommand{\R}{\mathbb R}
\newcommand{\Y}{\mathbb Y}
\newcommand{\Z}{\mathbb Z}
\newcommand{\Zp}{\mathbb Z_{\geq 0}}

\newcommand{\X}{\mathfrak{X}}
\newcommand{\Hc}{\mathcal{H}}

\DeclareMathOperator{\Conf}{Conf}

\DeclareMathOperator{\const}{const}
 \DeclareMathOperator{\sgn}{sgn}

\DeclareMathOperator{\Prob}{Prob} 
 
\DeclareMathOperator{\Pf}{Pf} \DeclareMathOperator{\Res}{Res}
\DeclareMathOperator{\diag}{diag}

\begin{document}
\title[Discrete ensembles]
 {\bf{Correlation Kernels for Discrete Symplectic and Orthogonal  Ensembles}}

\author{ Alexei Borodin}
\address{Department of Mathematics,  253-37,  Caltech, Pasadena, CA
91125} \email{borodin@caltech.edu}

\author{Eugene Strahov}
\address{Department of Mathematics, The Hebrew University of
Jerusalem, Givat Ram, Jerusalem
91904}\email{strahov@math.huji.ac.il}
\begin{abstract}
In \cite{widom} H. Widom derived formulae expressing  correlation
functions of orthogonal and symplectic ensembles of random
matrices in terms of orthogonal polynomials. We obtain similar
results for discrete ensembles with rational discrete logarithmic
derivative, and compute explicitly correlation kernels associated
to the classical Meixner and Charlier orthogonal polynomials.
\end{abstract}
\maketitle \tableofcontents
\section{Introduction}
The present paper addresses the following problem. Let $w(x)$ be a
classical discrete weight function, and consider discrete symplectic
and orthogonal ensembles associated to these weights, see Sections
\ref{MRDSE} and \ref{MRDOE} for precise definitions. By methods of
Random Matrix Theory (see e.g. Tracy and Widom \cite{tracy}) we find
$m$-point correlations in terms of Pfaffians with $2\times 2$ matrix
kernels. The problem is to express these kernels in terms of the
orthogonal polynomials associated with our weight function.

It is well known that Pfaffian expressions including $2\times 2$
matrix kernels appear in analysis of the orthogonal and the
symplectic ensembles of Random Matrix Theory, see, for example,
Mehta \cite{mehta}, Forrester \cite{forrester0}, Tracy and Widom
\cite{tracy}, Borodin and Strahov \cite{borodinstrahov}. Such
kernels also play a role in the works on the crossover between
matrix symmetries,  see Pandey and Mehta \cite{pandey1}, Mehta and
Pandey \cite{pandey2}, Nagao and Forrester \cite{nagao1},
superpositions of matrix ensembles, see Forrester and Rains
\cite{forresterrains, forrester2}, and also in the multi-matrix
models combining matrices of different symmetries, see Nagao
\cite{nagao2001, nagao2007}. Forrester and Nagao
\cite{forresternagao}, Borodin and Sinclair \cite{borodinsinclair}
give $2\times 2$ matrix kernels for ensembles of asymmetric real
matrices, in particular, for the eigenvalue statistics of Real
Ginibre Ensemble. Vicious random walkers, random involutions,
Pfaffian Schur process are examples of problems from combinatorics
and  statistical physics where Pfaffian formulas including
$2\times 2$ matrix kernels arise, see Nagao and Forrester
\cite{nagaoforrester2002}, Nagao, Katori and Tanemura
\cite{NagaoKatoriTanemura},  Nagao \cite{nagao2003}, Forrester,
Nagao and Rains \cite{forrester}, Borodin and Rains
\cite{borodin2}, Vuleti\'c \cite{vuletic}.

Often these kernels
 are constructed in terms of skew-orthogonal polynomials.
 Then a question arises how to compute
 the skew-orthogonal polynomials,
 and how to handle the Christoffel-Darboux  sums involving them.
 In some cases explicit formulas for
 skew-orthogonal polynomials  have been given in terms of related
 orthogonal polynomials, and
 matrix kernels and their asymptotic values have been computed.
This approach is developed in Nagao and Wadati \cite{nagaowadati},
Br$\acute{\mbox{e}}$zin and Neuberger \cite{brezinneuberger1991},
Adler, Forrester, Nagao and P. van Moerbeke \cite{adler}, see also
Forrester \cite{forrester0}, Chapter 5. Nagao \cite{nagao2007}
provides the list of the cases when the expressions of skew
orthogonal polynomials in terms of the classical orthogonal
polynomials are explicitly known, see Table 1.

Our definitions of discrete symplectic and orthogonal ensembles in
Sections \ref{MRDSE} and \ref{MRDOE} are motivated by relations
with $z$-measures on Young diagrams with the Jack parameter
$\theta=2$, as it is described in Section \ref{zmeasures}. For
ensembles obtained in Section \ref{zmeasures} skew-orthogonal
polynomials are not known, and we use  a discrete version of the
method developed by Widom in \cite{widom} in the context of
orthogonal and symplectic ensembles of Hermitian matrices. Widom
\cite{widom} gives general formulas expressing entries of $2\times
2$ matrix kernels in terms of the scalar kernels for the
corresponding unitary ensembles. Whenever the logarithmic
derivative of the weight in the definition of orthogonal or
symplectic ensemble under considerations is a rational function,
the entries of the $2\times 2$ matrix kernels are expressible in
terms of orthogonal polynomials, and are equal to the scalar
kernel plus extra terms. Similar results for ensembles with
Laguerre-type weights were obtained in physical literature, see
 Sener and  Verbaarschot \cite{sener}, Klein and Verbaarschot \cite{klein}.
 These papers show that the number of extra terms is
finite,  which leads to universality of  correlation kernels for
such ensembles.

 Formulae obtained in Widom \cite{widom}
are especially convenient for the asymptotic analysis since the
asymptotics of polynomials associated to rather general classes of
weights is known, see Deift,  Kriecherbauer,  McLaughlin,
Venakides, and Zhou \cite{deift1,deift2}, Bleher and Its
\cite{bleherits}. This enables one to use Widom's formulae in the
proof of the universality for the orthogonal and symplectic
ensembles, see Deift and Gioev \cite{deift3,deift4}, Deift, Gioev,
Kriecherbauer, and Vanlessen \cite{deift5}, Stojanovic
\cite{stojanovich}.
$$
$$
\begin{tabular}{|l|l|l|}
\hline   &     &   \\
 The weight & The orthogonality  & The family \\
       defining the families &  set    & of the orthogonal\\
                  &                    & polynomials\\
\hline             &                   &             \\
$w(x)=e^{-x^2}$ & $(-\infty,+\infty)$ & Hermite polynomials\\
                        &                     &  (see \cite{adler,brezinneuberger1991, nagaowadati})\\
\hline &                   &             \\
$w(x)=x^ae^{-x}$ & $(0,+\infty)$ & Laguerre polynomials\\
                        &                     &  (see \cite{adler,nagaoforrester1995})\\
\hline &                   &             \\
$w(x)=(1-x)^a(1-x)^b$ & $(-1,1)$ & Jacobi polynomials\\
                        &                     &  (see \cite{adler,nagaoforrester1995,nagaowadati})\\
\hline &                   &             \\
$w(x)=\left[\left(\frac{L}{2}+x\right)!\left(\frac{L}{2}-x\right)!\right]^{-2}$ & $\Z$ & Hahn polynomials\\
  $L$ is an even integer                      &                     &  (see \cite{nagaoforrester2002})\\
\hline &                   &             \\
$w(x)=1$ & $\{0,1,\ldots,L\}$ & Hahn polynomials\\
                        &                     &  (see \cite{nagao2007})\\
\hline &                   &             \\
$w(x)=q^x$       & $\Zp$ &   Meixner polynomials \\
                        &                     & $M_n(x;c=1,q)$ (see \cite{forrester})\\
\hline
\end{tabular}
$$
$$
Table 1. The cases in which skew-orthogonal polynomials are
explicitly known.\\
$$
$$

Our results for discrete symplectic and orthogonal ensembles are of
the similar kind as those obtained in Widom \cite{widom}. Whenever a
discrete analog of the logarithmic derivative of the weight is a
rational function the matrix kernels are expressible in terms of the
orthogonal polynomials associated to weight in the definition of the
ensemble. This is used to work out the cases of the Charlier
ensemble ($w(x)=\frac{a^x}{x!}$, $x\in\Zp$), and the Meixner
ensemble ($w(x)=\frac{(\beta)_x}{x!}c^x$, $x\in\Zp$). As an
application, we compute the continuous limit of our formulas
corresponding to the degeneration of the Meixner orthogonal
polynomials to the Laguerre orthogonal polynomials.
\\

\textbf{Acknowledgements.} We thank Percy Deift for discussions.
This work was supported in part by BSF grant BSF-2006333. The first
named author (A.B.) is grateful to Dimitri Gioev for helpful
discussion during early stages of this work. He was also partially
supported by the NSF grant DMS-0707163.

\section{Main results for discrete symplectic
ensembles}\label{MRDSE}
 Let $w(x)$ be a strictly positive real
valued function defined on $\Zp$ with finite moments, i.e. the
series $\sum_{x\in\Zp}w(x)x^{j}$ converges for all $j=0,1,\ldots$.
\begin{defn}\label{1DEF}
The $N$-point discrete symplectic ensemble with the weight
function $w$ and the phase space $\Zp$ is the random $N$-point
configuration in $\Zp$ such that the probability of a particular
configuration $x_1<\ldots <x_N$ is given by
$$
\Prob\left\{x_1,\ldots,x_N\right\}=Z_{N4}^{-1}\;\prod\limits_{i=1}^Nw(x_i)
\prod\limits_{1\leq i<j\leq N}(x_i-x_j)^2(x_i-x_j-1)(x_i-x_j+1).
$$
Here $Z_{N4}$ is a normalization constant which is assumed to be
finite.
\end{defn}
In what follows $Z_{N4}$ is referred to as the partition function
of the discrete symplectic ensemble under considerations.

Introduce a collection $\{P_n(\zeta)\}_{n=0}^{\infty}$ of complex
polynomials which is the collection of orthogonal polynomials
associated to the weight function $w$, and to the orthogonality
set $\Zp$. Thus
\begin{itemize}
    \item $P_n$ is a polynomial of degree $n$ for all
    $n=1,2,\ldots$, and $P_0\equiv\const$.
    \item If $m\neq n$, then
    $$\sum\limits_{x\in\Zp}P_m(x)P_n(x)w(x)=0.$$
\end{itemize}
For each $n=0,1,\ldots $ set
$\varphi_n(x)=\left(P_n,P_n\right)^{-1/2}_{w}P_n(x)w^{1/2}(x)$,
where $(.,.)_w$ denotes the following inner product on the space
$\C[\zeta]$ of all complex polynomials:
$$
\left(f(\zeta),g(\zeta)\right)_w:=\sum\limits_{x\in\Zp}f(x)g(x)w(x).
$$
We call $\varphi_n$ the normalized functions associated to the
orthogonal polynomials $P_n$.

Let $\Hc$ be the space spanned by the functions $\varphi_0,
\varphi_1,\ldots$.
\begin{defn}\label{DEFINITIONKOROP}
Suppose that there is a $2\times 2$ matrix valued kernel
$K_{N4}(x,y)$, $x, y\in\Zp$, such that for a general finitely
supported function $\eta$ defined on $\Zp$ we have
$$
Z_{N4}^{-1}\;\sum\limits_{(x_1<\ldots<x_N)\subset\Zp}\prod\limits_{i=1}^Nw(x_i)\left(1+\eta(x_i)\right)
\prod\limits_{1\leq i<j\leq N}(x_i-x_j)^2(x_i-x_j-1)(x_i-x_j+1)
$$
$$
=\sqrt{\det\left(I+\eta K_{N4}\right)},
$$
 where $K_{N4}$ is the operator
associated to the kernel $K_{N4}(x,y)$, and $\eta$ is  the
operator of multiplication by the function $\eta$.  $K_{N4}(x,y)$
is called the correlation kernel of the discrete symplectic
ensemble defined by the weight function $w(x)$ on the phase space
$\Zp$.
\end{defn}
An explanation of the term ``correlation kernel" can be found in
Tracy and Widom \cite{tracy}, $\S$2, 3.

 We introduce the operators $D_+, D_-$ and
$\epsilon$ which act on the elements of the space $\Hc$. The first
and the second operators, $D_+$ and $D_-$, are defined by the
expression:
$$
\left(D_{\pm}f\right)(x)=\sum\limits_{y\in\Zp}D_{\pm}(x,y)f(y),
$$
where the kernels $D_{\pm}(x,y)$ are given explicitly by
\begin{equation}\label{DPLUS(x,y)}
D_+(x,y)=\sqrt{\frac{w(x)}{w(x+1)}}\;\delta_{x+1,y},\;\;x,
y\in\Zp,
\end{equation}
\begin{equation}\label{DMINUS(x,y)}
D_-(x,y)=\sqrt{\frac{w(x-1)}{w(x)}}\;\delta_{x-1,y},\;\;x,
y\in\Zp.
\end{equation}
The third operator, $\epsilon$, is defined by the formula
\begin{equation}\label{OPERATOREPSILON}
\begin{split}
\left(\epsilon
\varphi\right)(2m)=-\sum\limits_{k=m}^{+\infty}\sqrt{\frac{w(2m)}{w(2k+1)}}
\frac{w(2m+1)w(2m+3)\ldots w(2k+1)}{w(2m)w(2m+2)\ldots w(2k)}\,\varphi(2k+1),\\
\left(\epsilon
\varphi\right)(2m+1)=\sum\limits_{k=0}^{m}\sqrt{\frac{w(2k)}{w(2m+1)}}
\frac{w(2k+1)w(2k+3)\ldots w(2m+1)}{w(2k)w(2k+2)\ldots
w(2m)}\,\varphi(2k),
\end{split}
\end{equation}
where $m=0,1,\ldots$.

 To make sure that $\epsilon \varphi$ is well defined for any
$\varphi\in\Hc$, we impose an additional assumption on the weight
function $w$: we assume that
$$
\frac{w(x-1)}{w(x)}=\frac{d_1(x)}{d_2(x)},\;\; x\geq 1,
$$
for some polynomials $d_1$ and $d_2$ such that
$\mbox{deg}\;d_1\geq \mbox{deg}\;d_2$ and if
$\mbox{deg}\;d_1=\mbox{deg}\;d_2$ then
$\underset{x\rightarrow\infty}{\lim}d_1(x)/d_2(x)>1$. This
implies, in particular, that $w(x-1)/w(x)>\const>1$ for $x\gg1$,
and one easily verifies that the series defining
$(\epsilon\varphi)(x)$ converges for any $\varphi\in\Hc$ and
$x\in\Z_{\geq 0}$.

Let us also introduce the operator $S_{N4}$ which acts in the same
space $\Hc$, and whose kernel is $S_{N4}(x,y)$. To write down
$S_{N4}(x,y)$ explicitly, introduce $2N\times 2N$ matrix $M^{(4)}$
whose $j,k$ entry $(j,k=0,1,\ldots ,2N-1)$ is
\begin{equation}\label{M4def}
M^{(4)}_{jk}=\sum\limits_{x\in\Zp}\varphi_j(x)\left(D\varphi_k\right)(x),
\end{equation}
where $D:=D_+-D_-$.
\begin{prop}\label{MP}
The matrix $M^{(4)}$ is invertible.
\end{prop}
All the proofs are delayed until Section
\ref{SectionDerKernelSympl}.

Write $(M^{(4)})^{-1}=(\mu_{jk}^{(4)})$, and define the kernel
$S_{N4}(x,y)$ by the formula:
\begin{equation}\label{KERNELS4N131}
S_{N4}(x,y)=\sum\limits_{j,k=0}^{2N-1}\varphi_j(x)\mu_{jk}^{(4)}\varphi_k(y),
\end{equation}
where $x, y\in\Zp$.
\begin{thm}\label{KN4THEOREM}The operator $K_{N4}$ (see Definition
\ref{DEFINITIONKOROP}) is expressible as
$$
K_{N4}=\left[\begin{array}{cc}
  D_+S_{N4} & -D_+S_{N4}D_- \\
  S_{N4} & -S_{N4}D_- \\
\end{array}\right].
$$
\end{thm}
\begin{rem}
As it is clear from the proof, the operator $K_{N4}$  for the
discrete symplectic ensemble can be also represented as
\begin{equation}\label{AlternativeFormulaKN4}
K_{N4}=\left[\begin{array}{cc}
  \nabla_+S_{N4} &  -\nabla_+S_{N4}\nabla_-\\
  S_{N4} & -S_{N4}\nabla_- \\
\end{array}\right].
\end{equation}
In the formula just written above the operators $\nabla_+,
\nabla_-$ are defined by
$$
\left(\nabla_+f\right)(x)=\sqrt{\frac{w(x)}{w(x+1)}}\left(f(x+1)-f(x)\right),
$$
$$
\left(\nabla_-f\right)(x)=\sqrt{\frac{w(x-1)}{w(x)}}\left(f(x)-f(x-1)\right).
$$
\end{rem}
Let $\Hc_N$ be the subspace of $\Hc$ spanned by the functions
$\varphi_0,\varphi_1,\ldots ,\varphi_{2N-1}$. Denote by $K_N$ the
projection operator onto $\Hc_N$. Its kernel is
$$
K_N(x,y)=\sum\limits_{k=0}^{2N-1}\varphi_k(x)\varphi_k(y).
$$
It is convenient to enlarge the domains of $D$ and $\epsilon$, and
to consider the operators
$$
D:\;\Hc+\epsilon\Hc\rightarrow\Hc+D\Hc,\\
$$
$$
\epsilon:\;\Hc+D\Hc\rightarrow\Hc+\epsilon\Hc.
$$
It is not hard to check that these operators are mutual inverse.
Denote by $D_{\Hc_N}$ the restriction of the operator $D$ to
$\Hc_N$.
\begin{thm}\label{MainTheorem}
The following operator identity holds true
$$
D_{\Hc_N}S_{N4}=\left(I_{\mathcal{H}_N+D\mathcal{H}_N}-[D,K_N]K_N\epsilon\right)^{-1}K_N.
$$
\end{thm}
The next Theorem gives the condition on the weight function $w(x)$
under which  the operator $S_{N4}$ can be written in an explicit
form.
\begin{thm}\label{THEOREM751}
Let $w(x)$ be a weight function such that for $x\geq 1$
$$
\frac{w(x-1)}{w(x)}=\frac{d_1(x)}{d_2(x)},
$$
where $d_1, d_2$ are polynomials of degree at most $m$ satisfying
the assumption above, and $d_1(0)=0$, $d_2(0)\neq 0$. Then
\begin{equation}
\left[D,K_N\right]K_N=\sum\limits_{i=1}^n\tilde{\psi}_i\otimes\psi_i,
\label{n}
\end{equation}
where $a\otimes b$ denotes the operator with the kernel
$a(x)b(y)$, $\psi_1,\ldots,\psi_n$ are elements of
$\mathcal{H}_N$, and $\tilde{\psi}_{1},\ldots ,\tilde{\psi}_{n}$
are elements of $\mathcal{H}^{\perp}_N$. Assume in addition that
the matrix $T_{ij}=\delta_{ij}+(\epsilon\psi_i,\tilde{\psi}_j)$,
$i,j=1,\ldots, n$ is invertible. Then
\begin{equation}
S_{N4}=\epsilon
K_N-\sum\limits_{i,j=1}^n(T^{-1})_{ij}(\epsilon\tilde{\psi}_i)\otimes\left(K_N\epsilon\psi_j\right).
\end{equation}
\end{thm}
Set $d_2(x)=\const\cdot (x-a_1)^{n_1}\ldots (x-a_l)^{n_l}$, and let
$n_{\infty}$ be the order of $\frac{w(x-1)}{w(x)}$ at $\infty$. As
will be clear from the proof of Theorem \ref{THEOREM751}, the number
$n$ from (\ref{n}) is bounded by
$n_{\infty}+\sum\limits_{i=1}^ln_{a_i}$. In Proposition
\ref{Propositionepsilonpsi1} we show that $n=1$ implies $T=1$.
\begin{cor}\label{COROLLARY1}
If the commutation relation between the operators $D$ and $K_N$
takes the form
$$
[D,K_N]=\lambda(\psi_1\otimes\psi_2+\psi_2\otimes\psi_1),
$$
where $\psi_1\in\mathcal{H}_N, \psi_2\in\mathcal{H^{\perp}}_N$,
and $\lambda$ is some constant, then
$$
S_{N4}=\epsilon K_N-\lambda\epsilon\psi_2\otimes\epsilon\psi_1.
$$
\end{cor}
The general formalism described above can be applied in particular
to the Meixner  and to the Charlier symplectic ensembles. The
weight function for the Meixner symplectic ensemble is by
definition the weight function associated to the Meixner
orthogonal polynomials:
\begin{equation}\label{MeinerWeight171}
w_{Meixner}(x)=\frac{(\beta)_x}{x!}c^x,\;\;x\in\Zp,
\end{equation}
where $\beta$ is a strictly positive real parameter, and $0<c<1$.
The weight function of the Charlier symplectic ensemble is defined
by
\begin{equation}\label{ChW172}
 w_{Charlier}(x)=\frac{a^x}{x!},\;\;x\in\Zp,
\end{equation}
where $a>0$. $w_{Charlier}$ is the weight function defining the
classical Charlier orthogonal polynomials\footnote{For definitions
and basic properties of the classical discrete orthogonal
polynomials see Ismail \cite{ismail}, and also Koekoek and
Swarttouw \cite{koekoek}}.

\begin{thm}\label{MeixnerChSYMPLECTICTHEOREM}
a) If $w(x)$ is the Meixner weight with the parameters $c$ and
$\beta$ defined by equation (\ref{MeinerWeight171}), then the
operator $S_{N4}$ whose kernel is defined by equation
(\ref{KERNELS4N131}) takes the following form:
\begin{equation}
S_{N4}=\epsilon
K_{N}+\frac{\sqrt{2N(2N+\beta-1)}}{(1-c)\sqrt{c}}\left(\epsilon\psi_2\right)\otimes\left(\epsilon\psi_1\right),
\nonumber
\end{equation}
where the operator $K_N$ has the kernel
\begin{equation}\label{formula2.9a}
K_N(x,y)=-\frac{\sqrt{2Nc(2N+\beta-1)}}{1-c}\frac{\varphi_{2N}(x)\varphi_{2N-1}(y)-\varphi_{2N-1}(x)\varphi_{2N}(y)}{x-y},
\end{equation}
the functions $\left\{\varphi_k(x)\right\}_{k=0}^{\infty}$ are the
normalized functions associated to the Meixner orthogonal
polynomials, the operator $\epsilon$ acts by the formula
\begin{equation}
\left(\epsilon \varphi\right)(2m)=
-\sqrt{c}\sum\limits_{l=0}^{+\infty}
\sqrt{\frac{(\frac{\beta}{2}+m)_{l+1}}{(\frac{\beta+1}{2}+m)_{l}}
\frac{(m+1)_{l}}{(m+\frac{1}{2})_{l+1}}}\,\varphi(2l+2m+1),
\nonumber
\end{equation}
\begin{equation}
\left(\epsilon \varphi\right)(2m+1)= \sqrt{c}\sum\limits_{l=0}^{m}
\sqrt{\frac{(-\frac{\beta}{2}-m)_{l+1}}{(-\frac{\beta-1}{2}-m)_{l}}
\frac{(-m)_{l}}{(-m-\frac{1}{2})_{l+1}}}\,\varphi(2m-2l), \nonumber
\end{equation}
where $m=0,1,\ldots$, and the functions $\psi_1$, $\psi_2$ are
defined by the expressions
\begin{equation}\label{psi1M}
\psi_1(x)=\sqrt{2Nc}\frac{\varphi_{2N}(x)}{x+\beta}-\sqrt{2N+\beta-1}\frac{\varphi_{2N-1}(x)}{x+\beta},
\end{equation}
\begin{equation}\label{psi2M}
\psi_2(x)=\sqrt{2N+\beta-1}\frac{\varphi_{2N}(x)}{x+\beta-1}-\sqrt{2Nc}\frac{\varphi_{2N-1}(x)}{x+\beta-1}.
\end{equation}

 b)If $w(x)$ is the Charlier weight with the
parameter $a$ (see equation (\ref{ChW172})), then the operator
$S_{N4}$ whose kernel is defined by equation (\ref{KERNELS4N131})
takes the following form:
\begin{equation}
S_{N4}=\epsilon
K_{N}+\sqrt{\frac{2N}{a}}\left(\epsilon\varphi_{2N}\right)\otimes\left(\epsilon\varphi_{2N-1}\right),
\nonumber
\end{equation}
where the operator $K_N$ has the kernel
\begin{equation}
K_N(x,y)=-\sqrt{2Na}\frac{\varphi_{2N}(x)\varphi_{2N-1}(y)-\varphi_{2N-1}(x)\varphi_{2N}(y)}{x-y},
\nonumber
\end{equation}
the functions $\left\{\varphi_k(x)\right\}_{k=0}^{\infty}$ are the
normalized functions associated to the Charlier orthogonal
polynomials, and  the operator $\epsilon$ acts as follows:
\begin{equation}
\left(\epsilon
\varphi\right)(2m)=-\sqrt{\frac{a}{2}}\sum\limits_{l=0}^{+\infty}
\frac{(m+1)_{l}}{(m+\frac{1}{2})_{l+1}}\,\varphi(2l+2m+1), \nonumber
\end{equation}
\begin{equation}
\left(\epsilon
\varphi\right)(2m+1)=\sqrt{\frac{a}{2}}\sum\limits_{l=0}^{m}
\frac{(-m)_{l}}{(-m-\frac{1}{2})_{l+1}}\,\varphi(2m-2l). \nonumber
\end{equation}
\end{thm}

\section{Main results for discrete orthogonal
ensembles}\label{MRDOE}
\begin{defn}\label{2DEF}
The $2N$-point discrete orthogonal ensemble with the weight
function $W$ and the phase space $\Zp$ is the random $2N$-point
configuration in $\Zp$ such that the probability of a particular
configuration $x_1<\ldots <x_{2N}$ is given by
\begin{equation}
\begin{split}
&\Prob\left\{x_1,\ldots,x_{2N}\right\}=\\
& \left\{
                                        \begin{array}{ll}
                                          Z_{N1}^{-1}\;
\prod\limits_{i=1}^{2N}W(x_i) \prod\limits_{1\leq
i<j\leq 2N}(x_j-x_i), & \hbox{if}\; x_i-x_{i-1}\; \hbox{is odd for any $i$, and $x_1$ is even}, \\
                                          0, & \hbox{otherwise.}
                                        \end{array}
                                      \right.
\end{split}
\nonumber
\end{equation}
Here $Z_{N1}$ is a normalization constant.
\end{defn}
In what follows we assume that the weight function $W(x)$ is such
that
\begin{equation}\label{RelWeights}
W(x-1)W(x)=w(x),\;\hbox{for}\;x\geq 1, \hbox{and}\;W(0)=w(0),
\end{equation}
where $w(x)$ is a strictly positive real valued function on $\Zp$
satisfying the same conditions as the weight function in the
definition of the discrete symplectic ensemble in Section
\ref{MRDSE}.

\begin{defn}\label{DefKorOrth}
Suppose that there is a $2\times 2$ matrix valued kernel
$K_{N1}(x,y)$, $x, y\in\Zp$, such that for an arbitrary finitely
supported function $\eta$ defined on $\Zp$ we have
$$
Z_{N1}^{-1}\;\sum\limits_{(x_1<\ldots<x_{2N})\subset\Zp}\prod\limits_{i=1}^{2N}
W(x_i)(1+\eta(x_i)) \prod\limits_{1\leq i<j\leq 2N}(x_j-x_i)
=\sqrt{\det\left(I+\eta K_{N1}\right)},
$$
where $K_{N1}$ is the operator associated with the kernel
$K_{N1}(x,y)$, and $\eta$ is the operator of multiplication by the
function $\eta$. Then the kernel of $K_{N1}$ is called the
correlation kernel of the discrete orthogonal ensemble defined by
the weight function $W(x)$ on the phase space $\Zp$.
\end{defn}
As in the symplectic case, details on the correlation functions
can be found in Tracy and Widom \cite{tracy}, \S 2, 3.

 Let $\epsilon$ and
$D$ be as in the previous section, and let $w(x)$ in the
definitions of these operators be given in terms of the weight
function $W(x)$ by formula (\ref{RelWeights}). Introduce the
operator $S_{N1}$ which acts in the same space $\Hc$, and whose
kernel is $S_{N1}(x,y)$. To write down $S_{N1}(x,y)$ explicitly
introduce $2N\times 2N$ matrix $M^{(1)}$ whose $j,k$ entry
$(j,k=0,1,\ldots ,2N-1)$ is
\begin{equation}\label{EntryM1}
M^{(1)}_{jk}=\sum\limits_{x,y\in\Zp}\epsilon(x,y)\varphi_j(x)\varphi_k(y).
\end{equation}
\begin{prop}\label{M1P}
The matrix $M^{(1)}$ is invertible.
\end{prop}
Write $(M^{(1)})^{-1}=(\mu_{jk}^{(1)})$, and define the kernel
$S_{N1}(x,y)$ by the formula:
\begin{equation}\label{KERNELS1N}
S_{N1}(x,y)=\sum\limits_{j,k=0}^{2N-1}\varphi_j(x)\mu_{jk}^{(1)}\varphi_k(y),
\end{equation}
where $x, y\in\Zp$.
\begin{thm}\label{TOOKOOOT}The operator $K_{N1}$ (see Definition
\ref{DefKorOrth}) is expressible as
$$
K_{N1}=\left[\begin{array}{cc}
  S_{N1}\epsilon & S_{N1} \\
  \epsilon S_{N1}\epsilon-\epsilon & \epsilon S_{N1} \\
\end{array}\right].
$$
\end{thm}
Denote by $\epsilon_{\Hc_N}$ the restriction of the operator
$\epsilon$ to $\Hc_N$. Recall that $K_N$ is the projection
operator on $\Hc_N$.
\begin{thm}\label{MainTheoremO}
The following operator identity holds true
$$
\epsilon_{\Hc_N}S_{N1}=\left(I_{\mathcal{H}_N+\epsilon\mathcal{H}_N}-[\epsilon,K_N]K_ND\right)^{-1}K_N.
$$
\end{thm}
\begin{thm}\label{MainTheoremOO}
Let $w(x)$ be as in Theorem \ref{THEOREM751}. Then
\begin{equation}\label{[e,KN]KN}
\left[\epsilon,K_N\right]K_N=\sum\limits_{i=1}^n\tilde{\eta}_i\otimes\eta_i,
\end{equation}
where $\eta_1,\ldots,\eta_n$ are elements of $\mathcal{H}_N$, and
$\tilde{\eta}_{1},\ldots ,\tilde{\eta}_{n}$ are elements of
$\mathcal{H}^{\perp}_N$. Assume in addition that the matrix
$U_{ij}=\delta_{ij}+(D\eta_i,\tilde{\eta}_j)$, $i,j=1,\ldots, n$,
is invertible. Then
\begin{equation}
S_{N1}=DK_N-\sum\limits_{i,j=1}^n(U^{-1})_{ij}(D\tilde{\eta}_i)\otimes\left(K_ND\eta_j\right).
\end{equation}
\end{thm}
The number of terms in (\ref{[e,KN]KN}) is bounded in the same way
as $n$ in Theorem \ref{THEOREM751}.
\begin{cor}\label{COROLLARY1ORTH}
If the commutation relation between the operators $D$ and $K_N$
takes the form
$$
[D,K_N]=\lambda(\psi_1\otimes\psi_2+\psi_2\otimes\psi_1),
$$
where $\psi_1\in\mathcal{H}_N, \psi_2\in\mathcal{H}^{\perp}_N$,
and $\lambda$ is some constant, then
$$
S_{N1}=D K_N-\lambda \psi_2\otimes\psi_1.
$$
\end{cor}
\begin{thm}\label{MeixnerChOOrtTheorem}
For Meixner or Charlier weight (see equations (\ref{ChW172}) and
(\ref{MeinerWeight171}) ), we have the following
expressions for the operator $S_{N1}$:\\
a) For the Meixner orthogonal ensemble
\begin{equation}
S_{N1}=DK_{N}+\frac{\sqrt{2N(2N+\beta-1)}}{(1-c)\sqrt{c}}\psi_2\otimes\psi_1,
\end{equation}
where the functions $\psi_1$ and $\psi_2$ are defined  by
equations (\ref{psi1M}) and (\ref{psi2M}) respectively.
 b)For the Charlier orthogonal ensemble
\begin{equation}
S_{N1}=DK_{N}+\frac{\sqrt{2N}}{a}\varphi_{2N}\otimes\varphi_{2N-1}.
\end{equation}
\end{thm}

\section{Discrete symplectic and orthogonal ensembles related with
$z$-measures}\label{zmeasures} Take $z,z'\in \mathbb{C}$,
$\theta>0$, $0<\xi< 1$, and define a distribution on the set of
all Young diagrams by
$$
M_{z,z',\theta,\xi}(\lambda)=(1-\xi)^t\xi^{|\lambda|}
\frac{(z)_{\lambda,\theta}(z')_{\lambda,\theta}}{H(\lambda,\theta)H'(\lambda,\theta)}.
$$
We have used the following notation: $t=zz'/\theta$;
$$
(z)_{\lambda,\theta}=\prod\limits_{(i,j)\in\lambda}(z+(j-1)-(i-1)\theta),
$$
where the product is taken over all boxes in a Young diagram
$\lambda$, $(i,j)$ stands for the box in $i$th row and $j$th
column; $|\lambda|$ is the number of boxes in $\lambda$;
$$
H(\lambda,\theta)=\prod\limits_{(i,j)\in\lambda}\left((\lambda_i-j)+(\lambda_j'-i)\theta+1\right),
$$
$$
H'(\lambda,\theta)=\prod\limits_{(i,j)\in\lambda}\left((\lambda_i-j)+(\lambda_j'-i)\theta+\theta\right),
$$
where $\lambda'$ denotes the transposed diagram. One can show that
$\sum_{\lambda\in \Y}M_{z,z',\theta,\xi}(\lambda)=1$, where the
sum is over the $\Y$ of all Young diagrams. If, for example,
$z'=\bar z$, then all the weights are nonnegative, and we obtain a
probability distribution on the set of all Young diagrams.
$M_{z,z',\theta,\xi}(\lambda)$ is called the $z$-measure. Details
and explanations of importance of $z$-measures in representation
theory can be found in Borodin and Olshanski \cite{BO}, Olshanski
\cite{olshanski2003}.
\begin{prop}
For $N=1,2,\ldots $ let $\Y(N)\subset\Y$ denote the set of
diagrams $\lambda$ with $l(\lambda)\leq N$. Under the bijection
between diagrams $\lambda\in\Y(N)$ and $N$-point configurations on
$\Zp$ defined by
$$
\lambda\longleftrightarrow x_{N-i+1}=\lambda_i-2i+2N\;\;
(i=1,\ldots, N)
$$
the $z$-measure with parameters $z=2N$, $\theta=2$,
$z'=2N+\beta-2$ turns into
$$
\Prob\left\{x_1,\ldots,x_N\right\}=\const\cdot\prod\limits_{i=1}^N\frac{(\beta)_{x_i}}{x_i!}\xi^{x_i}
\prod\limits_{1\leq i\leq j\leq
N}(x_i-x_j)^2(x_i-x_j-1)(x_i-x_j+1),
$$
which is precisely the discrete symplectic ensemble with the
Meixner weight in the sense of Definition \ref{1DEF}.
\end{prop}
\begin{proof}
The proof is  a  straightforward computation based on the
application of the explicit formulae for
$H(\lambda;2)H'(\lambda;2)$, see the proof of Lemma 3.5 in
\cite{BO}, and $(z)_{\lambda,\theta}$, see Section 1 in \cite{BO}.
\end{proof}
\begin{prop}
For $N=1,2,\ldots $ let $\Y'(2N)\subset\Y$ denote the set of
diagrams $\lambda$ with $l(\lambda')\leq 2N$. Under the bijection
between diagrams $\lambda\in\Y'(2N)$ and $2N$-point configurations
on $\Zp$ defined by
$$
\lambda\leftrightarrow x_{2N-i+1}=2\lambda_i'-i+2N\;\;
(i=1,\ldots, 2N)
$$
the $z$-measure with parameters $z=-2N$, $\theta=2$,
$z'=-2N-\beta$ turns into
$$
\Prob\left\{x_1,\ldots,x_{2N}\right\}=\const\cdot\prod\limits_{i=1}^{2N}\frac{[\beta]_{x_i}}{x_i!!}\;\xi^{\frac{x_i}{2}}
\prod\limits_{1\leq i\leq j\leq 2N}(x_j-x_i),
$$
where
$$
x_i!!=\left\{
        \begin{array}{ll}
          2\cdot 4\cdot \ldots \cdot x_i, & \hbox{$x_i$ is even,} \\
          1\cdot 3\cdot \ldots \cdot x_i, & \hbox{$x_i$ is odd,}
        \end{array}
      \right.
$$
and
$$
[\beta]_{x_i}=\left\{
        \begin{array}{ll}
          (x_i+\beta-1)(x_i+\beta-3)\ldots (\beta+1), & \hbox{$x_i$ is even,} \\
          (x_i+\beta-1) (x_i+\beta-3)\ldots \beta , & \hbox{$x_i$ is odd.}
        \end{array}
      \right.
$$
This  is  a discrete orthogonal ensemble  in the sense of
Definition \ref{2DEF}.
\end{prop}
\begin{proof}
The proof is  also a straightforward  computation based on the
formula
$$
\frac{1}{H(\lambda;2)H'(\lambda;2)}=\frac{\prod\limits_{1\leq
i<j\leq
l(\lambda')}(2\lambda_i'-i-2\lambda_j'+j)}{\prod_{i=1}^{l(\lambda')}
(2\lambda_i'-i+l(\lambda'))!}.
$$
\end{proof}
\section{The derivation of the correlation kernel for discrete
symplectic ensembles}\label{SectionDerKernelSympl} Recall that the
Pfaffian of a $2N\times 2N$ antisymmetric matrix $A=\parallel
A_{jk}\parallel_{j,k=1}^{2N}$ is defined as
$$
\Pf A=\underset{i_1<i_3<\ldots
<i_{2N-1}}{\underset{i_1<i_2,\ldots,i_{2N-1}<i_{2N}}{\sum\limits_{\sigma=(i_1,\ldots,i_{2N})\in
S_{2N}}}}\sgn(\sigma)A_{i_1i_2}\ldots A_{i_{2N-1}i_{2N}}.
$$
One has $(\Pf A)^2=\det A$.
\begin{lem}\label{LEMMADEBR}
Assume that $\varphi_1,\ldots,\varphi_{2N}$ and
$\psi_1,\ldots,\psi_{2N}$ are arbitrary finitely supported
functions on $\Zp$. Set
$$
\underline{\varphi}(.)=\left[\begin{array}{c}
                          \varphi_1(.) \\
                          \vdots \\
                          \varphi_{2N}(.)
                        \end{array}\right],\;\;
\underline{\psi}(.)=\left[\begin{array}{c}
                          \psi_1(.) \\
                          \vdots \\
                          \psi_{2N}(.)
                        \end{array}\right],
$$
and introduce a $2N\times 2N$ antisymmetric matrix
$A=\left[A_{ij}\right]_{i,j=1}^{2N}$ whose entries, $A_{ij}$, are
given by
$$
A_{ij}=\sum\limits_{x\in\Zp}\left[\varphi_i(x)\psi_j(x)-\psi_i(x)\varphi_j(x)\right].
$$
 We
have
\begin{equation}\label{DdeBruijin}
\sum\limits_{\underline{x}=(x_1<\ldots<x_N)\subset\Zp}
\det\left[\underline{\varphi}(x_1),\underline{\psi}(x_1),
\ldots,\underline{\varphi}(x_N),\underline{\psi}(x_N)\right]=\Pf
A.
\end{equation}
\end{lem}
\begin{proof}
This is one of de Bruijn's formulas, see de Bruijn
\cite{deBruijn}.
\end{proof}
\begin{lem}\label{LEMMAPOLYNOMY}
Let
$$
\pi_{i-1}(x)=x^{i-1}+\ldots,\;\;\;i=1,\ldots ,2N,
$$
is an arbitrary system of monic polynomials of degrees $0,\ldots,
2N-1$. Set
$$
\varphi_i(x)=\pi_{i-1}(x),\;\;\psi_i(x)=\pi_{i-1}(x+1).
$$
Then
$$
\prod\limits_{1\leq i<j\leq
N}\left(x_i-x_j\right)^2\left(\left(x_i-x_j\right)^2-1\right) =
\det\left[\underline{\varphi}(x_1),\underline{\psi}(x_1),
\ldots,\underline{\varphi}(x_N),\underline{\psi}(x_N)\right].
$$
\end{lem}
\begin{proof}
We have
\begin{equation}
\begin{split}
\prod\limits_{1\leq i<j\leq
N}&\left(x_i-x_j\right)^2\left(\left(x_i-x_j\right)^2-1\right) \\
&=\prod\limits_{1\leq i<j\leq
N}\left(x_i-x_j\right)^2\left(x_i-x_j+1\right)\left(x_i-x_j-1\right)\\
&=V(x_1+1,x_1,x_2+1,x_2,\ldots,x_N+1,x_N)\\
&=(-1)^NV(x_1,x_1+1,x_2,x_2+1,\ldots,x_N,x_N+1)\\
&=(-1)^N(-1)^{\frac{2N(2N-1)}{2}}\det\left(\pi_{i-1}(x_1),\pi_{i-1}(x_1+1),\ldots,\pi_{i-1}(x_N),\pi_{i-1}(x_N+1)\right)\\
&=\det\left[\underline{\varphi}(x_1),\underline{\psi}(x_1),
\ldots,\underline{\varphi}(x_N),\underline{\psi}(x_N)\right].
\end{split}
\nonumber
\end{equation}
\end{proof}
\textbf{Proof of Proposition \ref{MP}.}\\
Applying Lemma \ref{LEMMADEBR} and Lemma \ref{LEMMAPOLYNOMY} we
obtain the identity
$$
Z_{N4}=\Pf\left[Q_{ij}\right]_{i,j=0}^{2N-1},
$$
where
\begin{equation}\label{MatrixQ}
Q_{ij}=\sum\limits_{x\in\Zp}w(x)\left(\pi_i(x)\pi_j(x+1)-\pi_i(x+1)\pi_j(x)\right),
\end{equation}
and $0\leq i,j\leq 2N-1$. Since the partition function $Z_{N4}$ of
the discrete symplectic ensemble under considerations is strictly
positive, the determinant of the matrix $Q$ is nonzero, and the
matrix $Q$ is non-degenerate. The matrix $M^{(4)}$ defined by
equation (\ref{M4def}) is related to $Q$ via
$Q=\diag(\|\pi_1\|^{1/2},\ldots ,\|\pi_{2N-1}\|^{1/2})\cdot
M\cdot\diag(\|\pi_1\|^{1/2},\ldots,\|\pi_{2N-1}\|^{1/2})$.\qed\\
\textbf{Proof of Theorem \ref{KN4THEOREM}}.\\
 Let $\zeta(x)$ be some finitely supported function defined on
$\Zp$. The following identity holds true
\begin{equation}\label{Azeta1}
\begin{split}
&\sum\limits_{(x_1<x_2<\ldots<x_N)\subset\Zp}\prod\limits_{i=1}^Nw(x_i)\zeta(x_i)
\prod\limits_{1\leq i<j\leq
N}(x_i-x_j)^2(x_i-x_j-1)(x_i-x_j+1)=\Pf\left[A_{ij}(\zeta)\right]_{i,j=1}^{2N},
\end{split}
\end{equation}
where the matrix elements $A_{ij}(\zeta)$ are defined by the
formula
\begin{equation}\label{Azeta2}
A_{ij}(\zeta)=
\sum\limits_{x\in\Zp}\left[\pi_{i-1}(x)\pi_{j-1}(x+1)-\pi_{i-1}(x+1)\pi_{j-1}(x)\right]w(x)\zeta(x).
\end{equation}
By the same argument as in Tracy and Widom \cite{tracy}, \S 8, we
find
\begin{equation}
\begin{split}
&K_{N4}(x,y)=\\
&\left[\begin{array}{cc}
               \sum\limits_{i,j=0}^{2N-1}
               \pi_i(x+1)Q^{-1}_{ij}\pi_j(y)w^{1/2}(x)
               w^{1/2}(y) & -\sum\limits_{i,j=0}^{2N-1}\pi_i(x+1)Q^{-1}_{ij}\pi_j(y+1)w^{1/2}(x)w^{1/2}(y) \\
               \sum\limits_{i,j=0}^{2N-1}
               \pi_i(x)Q^{-1}_{ij}\pi_j(y)w^{1/2}(x)
               w^{1/2}(y) & -\sum\limits_{i,j=0}^{2N-1}\pi_i(x)Q^{-1}_{ij}\pi_j(y+1)w^{1/2}(x)w^{1/2}(y)
             \end{array}
\right],
\end{split}
 \nonumber
\end{equation}
where the variables $x,y$ take values in $\Zp$, $Q_{ij}$ is
defined by equation (\ref{MatrixQ}), and
$\pi_i(x)=x^{i}+\ldots;\;\;i=0,1 \ldots $ is an arbitrary system
of monic polynomials of the discrete variable $x$. It is now
straightforward  to check that the kernel $K_{N4}(x,y)$ just
written above is exactly the kernel of the operator $K_{N4}$,
where
$$
K_{N4}=\left[\begin{array}{cc}
  D_+S_{N4} & -D_+S_{N4}D_- \\
  S_{N4} & -S_{N4}D_- \\
\end{array}\right].
$$
Finally observe that the matrix $A_{ij}(\zeta)$ remains unchanged
if we replace $\pi_{j-1}(x+1)$ by
$\pi_{j-1}(x+1)-\sqrt{\frac{w(x)}{w(x+1)}}\pi_{j-1}(x)$, and
$\pi_{i-1}(x+1)$ by
$\pi_{i-1}(x+1)-\sqrt{\frac{w(x)}{w(x+1)}}\pi_{i-1}(x)$ in
equation (\ref{Azeta2}). This results in formula
(\ref{AlternativeFormulaKN4}). Theorem \ref{KN4THEOREM} is
proved.\qed
\section{The derivation of the correlation kernel for
discrete orthogonal ensembles}
\begin{lem}\label{LemmaPEREpis'}
The probability of a particular configuration $x_1<\ldots<x_{2N}$
of the discrete orthogonal ensemble (see Definition \ref{2DEF})
can be rewritten as
\begin{equation}\label{NewDef}
\Prob\left\{x_1,\ldots,x_{2N}\right\}=\tilde{Z}_{N1}^{-1}\prod\limits_{i=1}^{2N}w^{1/2}(x_i)\prod\limits_{1\leq
i<j\leq 2N}(x_i-x_j)\Pf[\epsilon(x_i,x_j)]_{i,j=1}^{2N}
\end{equation}
where $w(x)$ is defined in terms of the weight function $W(x)$ by
formula (\ref{RelWeights}).
\end{lem}
\begin{proof}
We will compute the Pfaffian in equation (\ref{NewDef}), and will
show  that (\ref{NewDef}) coincides with the expression for the
probability of a particular configuration $x_1<\ldots<x_{2N}$ in
Definition \ref{2DEF}. Observe that the semi-infinite matrix
$\epsilon$  defined by equation (\ref{OPERATOREPSILON}) is
representable as follows
$$
\epsilon=\mathcal{F}\Upsilon\mathcal{F},
$$
where
$$
\mathcal{F}=\left[\begin{array}{ccccc}
                 f(0) & 0 & 0 & 0 & \ldots  \\
                 0 & f(1) & 0 & 0 & \ldots  \\
                 0 & 0 & f(2) & 0 & \ldots \\
                 0 & 0 & 0 & f(3) & \ldots \\
                 \vdots & \vdots & \vdots & \vdots & \ddots
               \end{array}
\right],
$$
$f(0)$, $f(1)$, $f(2)$, $\ldots$ are defined for $k=0,1,2,\ldots $
by
$$
f(2k)=\frac{1}{\sqrt{w(2k)}}\frac{w(2)w(4)\ldots
w(2k)}{w(1)w(3)\ldots
w(2k-1)},\;\;f(2k+1)=\frac{1}{\sqrt{w(2k+1)}}\frac{w(1)w(3)\ldots
w(2k+1)}{w(2)w(4)\ldots w(2k)},
$$
and
$$
\Upsilon=\left[\begin{array}{ccccccccc}
                 0 & -1 & 0 & -1 & 0 & -1 & 0 & -1 &\ldots  \\
                 1 & 0 & 0 & 0 & 0 & 0 & 0 & 0 & \ldots  \\
                 0 & 0 & 0 & -1 & 0 & -1 & 0 & -1 &  \ldots \\
                 1 & 0 & 1 & 0 & 0 & 0 & 0 & 0 & \ldots  \\
                 0 & 0 & 0 & 0 & 0 & -1 & 0 & -1 &  \ldots \\
                 1 & 0 & 1 & 0 & 1 & 0 & 0 & 0 & \ldots  \\
                 \vdots & \vdots & \vdots & \vdots & \vdots & \vdots & \vdots & \vdots & \ddots
               \end{array}
\right].
$$
That is, $\Upsilon$ is an antisymmetric matrix whose entries are
defined by the relations
$$
\Upsilon(2i+1,2j+1)=\Upsilon(2i,2j)=0,\;\hbox{for any $i,j\geq
0$},
$$
$$
\Upsilon(2i+1,2j+2)=0, \;\hbox{for $0\leq i\leq j$},
$$
$$
\Upsilon(2i,2j+1)=-1, \;\hbox{for $0\leq i\leq j$}.
$$
  Since
relation (\ref{RelWeights}) implies that
$$
W(x)=\left\{
       \begin{array}{ll}
         \frac{w(1)w(3)\ldots w(x)}{w(2)w(4)\ldots w(x-1)}, & x\;\hbox{is odd}, \\
         \frac{w(2)w(4)\ldots w(x)}{w(1)w(3)\ldots w(x-1)}, &
x\;\hbox{is even},
       \end{array}
     \right.
$$
we obtain
\begin{equation}
\begin{split}
&\prod\limits_{i=1}^{2N}w^{1/2}(x_i)\Pf[\epsilon(x_i,x_j)]_{i,j=1}^{2N}\\
&=\left(\prod\limits_{i=1}^{2N}w^{1/2}(x_i)\right)\left(\prod\limits_{i=1}^{2N}f(x_i)\right)
\Pf\left[\Upsilon(x_1,\ldots,x_{2N}\biggl\vert x_1,\ldots,x_{2N})\right]\\
&=\left(\prod\limits_{i=1}^{2N}W(x_i)\right)
\Pf\left[\Upsilon(x_1,\ldots,x_{2N}\biggl\vert
x_1,\ldots,x_{2N})\right].
\end{split}
\nonumber
\end{equation}
The proof is completed by the following Lemma, cf. Borodin and
Strahov \cite{borodinstrahov}, Section 3.2.
\end{proof}
\begin{prop}
$$
\Pf\left[\Upsilon(x_1,\ldots,x_{2N}\biggl\vert
x_1,\ldots,x_{2N})\right]= \left\{
  \begin{array}{ll}
    (-1)^N, & \hbox{if}\; x_i-x_{i-1}\;\hbox{is odd, and $x_1$ is even},  \\
    0, & \hbox{otherwise.}
  \end{array}
\right.
$$
\end{prop}
\begin{proof}
If $x_1$ is odd, then the first row of the matrix
$\Upsilon(x_1,\ldots,x_{2N}\biggl\vert x_1,\ldots,x_{2N})$
consists of only zeros. Thus, if
$\Pf\left[\Upsilon(x_1,\ldots,x_{2N}\biggl\vert
x_1,\ldots,x_{2N})\right]\neq 0$, $x_1$ must be even. Now assume
that $x_{2i-1}$ and $x_{2i}$ have the same parity. In this case
rows $2i-1$ and $2i$ of $\Upsilon(x_1,\ldots,x_{2N}\biggl\vert
x_1,\ldots,x_{2N})$ are equal to each other. Therefore, if
$\Pf\left[\Upsilon(x_1,\ldots,x_{2N}\biggl\vert
x_1,\ldots,x_{2N})\right]\neq 0$, the elements of the set
$X=(x_1,\ldots, x_{2N})$ are such that $x_1$ is even, $x_2$ is
odd, $x_3$ is even, and so on. This proves the condition on the
parity for the configurations $X=(x_1,\ldots, x_{2N})$, for which
$\Pf\left[\Upsilon(x_1,\ldots,x_{2N}\biggl\vert
x_1,\ldots,x_{2N})\right]\neq 0$. Moreover, using the definition
of Pfaffian, it is not hard to conclude that
$\Pf\left[\Upsilon(x_1,\ldots,x_{2N}\biggl\vert
x_1,\ldots,x_{2N})\right]=(-1)^N$ for the configurations for which
$\Pf\left[\Upsilon(x_1,\ldots,x_{2N}\biggl\vert
x_1,\ldots,x_{2N})\right]\neq 0$.
\end{proof}

\begin{lem}\label{DeBruijnOrt}
If $\phi_1,\ldots,\phi_{2N}$ are arbitrary finitely supported
functions on $\Zp$, and $\epsilon(x,y)$ is an antisymmetric
function defined on $\Zp$, then
\begin{equation}\label{711OrtLem}
\sum\limits_{x_1<\ldots
<x_{2N}\subset\Zp}\det\left(\phi_j(x_k)\right)_{j,k=1}^{2N}\Pf\left[\epsilon(x_i,x_j)\right]_{i,j=1}^{2N}
=\Pf\left[\sum\limits_{x,y\in\Zp}\epsilon(x,y)\phi_j(x)\phi_k(y)\right].
\end{equation}
\end{lem}
\begin{proof}
This is another de Bruijn formula, see \cite{deBruijn}.
\end{proof}
\textbf{Proof of Proposition \ref{M1P}}\\
Consider the partition function $\tilde{Z}_{N1}$, which is defined
by
$$
\tilde{Z}_{N1}=\sum\limits_{x_1<\ldots<
x_{2N}}\prod\limits_{i=1}^{2N}w^{1/2}(x_i)\prod\limits_{1\leq
i<j\leq 2N}(x_i-x_j)\Pf[\epsilon(x_i,x_j)]_{i,j=1}^{2N}.
$$
Write
$$
\prod\limits_{i=1}^{2N}w^{1/2}(x_i)\prod\limits_{1\leq i<j\leq
2N}(x_i-x_j)=(-1)^N\det\left[w^{1/2}(x_k)\pi_j(x_k)\right],
$$
where $0\leq j\leq 2N-1$,  $1\leq k\leq 2N$, and
$\left\{\pi_j(x)\right\}$ is an arbitrary system of monic
polynomials. Apply Lemma \ref{DeBruijnOrt} and obtain
$$
\tilde{Z}_{N1}=(-1)^N\det\left[\sum\limits_{x,y\in\Zp}\epsilon(x,y)\pi_j(x)w^{1/2}(x)\pi_k(y)w^{1/2}(y)\right]_{k,j=0}^{2N-1}.
$$
Since $\tilde{Z}_{N1}$ is nonzero, the matrix with the $j, k$
entry
$$
\sum\limits_{x,y\in\Zp}\epsilon(x,y)\pi_j(x)w^{1/2}(x)\pi_k(y)w^{1/2}(y)
$$
is non-degenerate. This implies that the matrix $M^{(1)}$ whose
$j, k$ entry is given by formula (\ref{EntryM1}) is non-degenerate
as well.\qed

\textbf{Proof of Theorem \ref{TOOKOOOT}}\\
Use Lemma \ref{LemmaPEREpis'} to rewrite the formula in definition
\ref{DefKorOrth} as follows
$$
\tilde{Z}_{N1}^{-1}\;\sum\limits_{x_1<\ldots<x_{2N}}\prod\limits_{i=1}^{2N}
w^{1/2}(x_i)(1+\eta(x_i)) \prod\limits_{1\leq i<j\leq
2N}(x_i-x_j)\Pf\left[\epsilon(x_i,x_j)\right]_{i,j=1}^{2N}
=\sqrt{\det\left(I+\eta K_{N1}\right)}.
$$
The rest of the argument is very similar to the derivation of the
correlation kernel in $\S 9$ of Tracy and Widom \cite{tracy}, and
we omit the details. \qed

\section{The general identities}
\begin{lem}\label{LEMMA1}
We have for $0\leq i\leq 2N-1$
\begin{equation}\label{Lemma1Equation1}
\left(S_{N4}K_ND\varphi_i\right)(x)=\varphi_i(x),\;\;\left(S_{N1}K_N\epsilon\varphi_i\right)(x)=\varphi_i(x).
\end{equation}
\begin{equation}\label{Lemma1Equation2}
S_{N4}\mid_{\mathcal{H}^{\perp}_N}=0,\;\;S_{N1}\mid_{\mathcal{H}^{\perp}_N}=0.
\end{equation}
where $\Hc^{\perp}_N$ denotes the complement of $\Hc_N$ in $\Hc$.
\end{lem}

\begin{proof}
a)For $0\leq i\leq 2N-1$, and $x\in\Zp$ we have
\begin{equation}
\begin{split}
\left(K_ND\varphi_i\right)(x)&=
\sum\limits_{j=0}^{2N-1}\varphi_j(x)\sum\limits_{y\in\Zp}\varphi_{j}(y)\left(D\varphi_i\right)(y).
\end{split}
\nonumber
\end{equation}
The second sum in the equation above can be rewritten as follows
\begin{equation}
\begin{split}
\sum\limits_{y\in\Zp}\varphi_{j}(y)\left(D\varphi_i\right)(y)
&=\sum\limits_{y\in\Zp}\varphi_{j}(y)\left(\sqrt{\frac{w(y)}{w(y+1)}}\varphi_i(y+1)
-\sqrt{\frac{w(y-1)}{w(y)}}\varphi_i(y-1)\right)\\
&=\sum\limits_{y\in\Zp}p_j(y)p_i(y+1)w(y)-\sum\limits_{y\in\Zp}p_j(y)p_i(y-1)w(y-1)\\
&=\sum\limits_{y\in\Zp}\left(p_j(y)p_i(y+1)-p_j(y+1)p_i(y)\right)w(y)\\
&=M_{ji}^{(4)},
\end{split}
\nonumber
\end{equation}
where $\{p_j\}$ is the family of polynomials orthonormal with
respect to the weight $w$. Therefore,
$$
\left(K_ND\varphi_i\right)(x)=\sum\limits_{j=0}^{2N-1}\varphi_j(x)M_{ji}^{(4)}.
$$
The action by the operator $S_{N4}$ from the left gives
\begin{equation}
\begin{split}
\left(S_{N4}K_ND\varphi_i\right)(x)&=
\sum\limits_{y\in\Zp}S_{N4}(x,y)\sum\limits_{j=0}^{2N-1}\varphi_{j}(y)M_{ji}^{(4)}\\
&=\sum\limits_{y\in\Zp}\sum\limits_{k,l=0}^{2N-1}\varphi_{k}(x)\mu_{kl}^{(4)}\varphi_{l}(y)\sum\limits_{j=0}^{2N-1}
\varphi_{j}(y)M_{ji}^{(4)} =\varphi_i(x). \nonumber
\end{split}
\end{equation}
b) It is clear from the  definition of the operator $S_{N4}$ that
$\left(S_{N4}\varphi_j\right)(x)=0$ for $j=2N, 2N+1,\ldots .$ In
other words, $S_{N4}\mid_{\mathcal{H}^{\perp}_N}=0.$ \\
c) The formulas for $S_{N1}$ can be proved by the same procedure,
replacing $D$ everywhere by $\epsilon$.
\end{proof}

Lemma \ref{LEMMA1} shows that the operators $S_{N4}$, $S_{N1}$,
$K_{N}$, $D$ and $\epsilon$ satisfy the same algebraic relations
as the corresponding operators in the continuous case considered
in Widom \cite{widom}, Sections 2-4. Therefore Theorem
\ref{MainTheorem} and Theorem \ref{MainTheoremO} can be
established by arguments from Widom \cite{widom}, Sections 2-4.
Here we only give a simple "matrix" explanation why formulae in
Theorem \ref{MainTheorem} and Theorem \ref{MainTheoremO} indeed
hold true.

Let us write each operator in the basis $\varphi_0,
\varphi_1,\varphi_2,\ldots$. Then each operator $A$ which acts in
$\Hc$ has the representation
$$
A=\left(\begin{array}{cc}
          (A)_N & (A)_1 \\
          (A)_0 & (A)_2
        \end{array}
\right)
$$
where the upper-left corner corresponds to the subspace $\Hc_N$
spanned by $\varphi_0,\ldots,\varphi_{2N-1}$. In particular, $K_N$
is representable as
$$
K_N=\left(\begin{array}{cc}
          I_N & 0 \\
          0 & 0
        \end{array}
\right),
$$
and we have
$$
(A)_{\Hc_N}=A\cdot K-N=\left(\begin{array}{cc}
          (A)_N & 0 \\
          (A)_0 & 0
        \end{array}
\right),\;\;(A)_{\Hc_N^{\bot}}=A\cdot
(I-K_N)=\left(\begin{array}{cc}
          0 & (A)_1 \\
          0 & (A)_2
        \end{array}
\right).
$$
In what follows we formally manipulate with operators. Equation
(\ref{Lemma1Equation1}) implies
$$
(K_NS_{N4}K_N)(K_NDK_N)=K_N.
$$
We can rewrite the above identity as follows
$$
\left(\begin{array}{cc}
          (S_{N4})_N & 0 \\
          0 & 0
        \end{array}
\right)\left(\begin{array}{cc}
          (D)_N & 0 \\
          0 & 0
        \end{array}
\right)=\left(\begin{array}{cc}
          I_N & 0 \\
          0 & 0
        \end{array}
\right).
$$
Therefore $(S_{N4})_{\Hc_N}=\left(\begin{array}{cc}
          (D)_N^{-1} & 0 \\
          0 & 0
        \end{array}
\right)$, and we obtain
$$
(D)_{\Hc_N}(S_{N4})_{\Hc_N}=\left(\begin{array}{cc}
          I_N & 0 \\
          (D)_0(D)_N^{-1} & 0
        \end{array}
\right).
$$
In order to express $(D)_0(D)_N^{-1}$  use the relation
$$
\epsilon D=I.
$$
This relation can be rewritten as
$$
\left(\begin{array}{cc}
          (\epsilon)_N & (\epsilon)_1 \\
          (\epsilon)_0 & (\epsilon)_2
        \end{array}
\right)\left(\begin{array}{cc}
          (D)_N & (D)_1 \\
          (D)_0 & (D)_2
        \end{array}
\right)=\left(\begin{array}{cc}
          I_N & 0 \\
          0 & I
        \end{array}
\right),
$$
and we obtain $(\epsilon)_N(D)_N=I_N-(\epsilon)_1(D)_0$.
Multiplying this equation by $(D)_0$ from the left we have
$$
(D)_0(\epsilon)_N(D)_N=(D)_0(I_N-(\epsilon)_1(D)_0)=(I_N-(D)_0(\epsilon)_1)(D)_0,
$$
therefore
$$
(D)_0(D)_N^{-1}=(I-(D)_0(\epsilon)_1)^{-1}(D)_0(\epsilon)_N.
$$
We conclude that the following formula holds
$$
(D)_{\Hc_N}(S_{N4})_{\Hc_N}=\left(\begin{array}{cc}
          I_N & 0 \\
          (I-(D)_0(\epsilon)_1)^{-1}(D)_0(\epsilon)_N & 0
        \end{array}
\right).
$$
Now we are going to show that
\begin{equation}\label{explanationseven}
 \left(\begin{array}{cc}
          I_N & 0 \\
          (I-(D)_0(\epsilon)_1)^{-1}(D)_0(\epsilon)_N & 0
        \end{array}
\right)=\left(I-(DK_N-K_NDK_N)\epsilon\right)^{-1}K_N.
\end{equation}
Indeed,
$$
DK_N-K_NDK_N=\left(\begin{array}{cc}
          0 & 0 \\
          (D)_0 & 0
        \end{array}
\right),
$$
$$
I-(DK_N-K_NDK_N)\epsilon=\left(\begin{array}{cc}
          0 & 0 \\
          -(D)_0(\epsilon)_N & I-(D)_0(\epsilon)_1
        \end{array}
\right),
$$
and using the formula $\left(\begin{array}{cc}
          1 & 0 \\
          a & b
        \end{array}
\right)^{-1}=\left(\begin{array}{cc}
          1 & 0 \\
          -b^{-1}a & b^{-1}
        \end{array}
\right)$ we get equation $(\ref{explanationseven})$. Thus we have
shown that
$$
(D)_{\Hc_N}(S_{N4})_{\Hc_N}=\left[I-(DK_N-K_NDK_N)\epsilon\right]^{-1}K_N.
$$
Taking into account equation (\ref{Lemma1Equation2}), we obtain
the formula in Theorem \ref{MainTheorem}. The formula in Theorem
\ref{MainTheoremO} can be deduced by similar computations.
\section{Proof of Theorem \ref{THEOREM751}.}
The idea of the proof is the same as in Widom \cite{widom1},
section 3.

Recall that the operator $D$ acts on the elements of $\Hc_N$
according to the formula
$$
(D\varphi)(x)=\sqrt{\frac{w(x)}{w(x-1)}}\varphi(x+1)-\sqrt{\frac{w(x-1)}{w(x)}}\varphi(x-1).
$$
Set $L_D=(I-K_N)D$, and agree that the domain of this operator is
$\Hc_N$. Denote by $\mathcal{N}_{D}$ the null space of $L_D$, i.e.
$$
\mathcal{N}_D=\{\varphi\vert L_D\varphi=0, \varphi\in\Hc_N\}.
$$
We want to find $\mathcal{N}_D^{\perp}$.

Let us find the null space $\mathcal{N}_{D}$ of $L_D=(I-K_N)D$.
The general element of $\Hc_N$ is of the form $\varphi=pw^{1/2}$,
where $\deg p\leq 2N-1$. Such $\varphi$ is an element of
$\mathcal{N}_D$ if and only if $D\varphi$ is an element of
$\Hc_N$. We have
$$
(D\varphi)(x)=\sqrt{w(x)}\left[p(x+1)-\frac{w(x-1)}{w(x)}p(x-1)\right].
$$
It follows that $\varphi$ is an element of $\mathcal{N}_D$ if and
only if  $\frac{w(x-1)}{w(x)}p(x-1)$ is a polynomial of degree
less or equal to $2N-1$.

 We will denote by $a_1, a_2,\ldots, a_l$   finite poles of $\frac{w(x-1)}{w(x)}$.
 Let $(\Hc_N)_{a_i}$
be the subspace of those $\varphi$, for which
$\frac{w(x-1)}{w(x)}p(x-1)=O(1)$ in a neighborhood of $a_i$, and
let $(\Hc_N)_{\infty}$ be the subspace of those $\varphi$, for
which $\frac{w(x-1)}{w(x)}p(x-1)=O(x^{2N-1})$ as
$x\rightarrow\infty$. Then
$$
\mathcal{N}_{D}=\left(\bigcap\limits_{i=1}^l(\Hc_N)_{a_i}\right)\bigcap(\Hc_N)_{\infty},
$$
and
$$
\mathcal{N}_{D}^{\perp}=\left(\sum\limits_{i=1}^l(\Hc_N)_{a_i}^{\perp}\right)+(\Hc_N)_{\infty}^{\perp}.
$$
Let $a_i$ be the finite pole of $\frac{w(x-1)}{w(x)}$ of order
$n_i$.  Observe that $p(x-1)\frac{w(x-1)}{w(x)}=O(1)$ if and only
if
\begin{equation}\label{pppcondition}
p(a_i-1)=0,\; p'(a_i-1)=0,\; \ldots,\; p^{(n_i-1)}(a_i-1)=0.
\end{equation}
Therefore $\varphi$ is an element of $(\Hc_N)_{a_i}$ if and only
if condition (\ref{pppcondition}) is satisfied.

 Now expand
$\varphi$ in terms of basis elements $\varphi_1$, $\ldots$,
$\varphi_{2N-1}$ of $\Hc_N$
$$
\varphi=\sum\limits_{j=0}^{2N-1}B_j\varphi_j.
$$
Thus
$$
p(x)=\sum\limits_{j=0}^{2N-1}B_jp_j(x),
$$
and condition (\ref{pppcondition}) implies that $\varphi$ is an
element of $(\Hc_N)_{a_i}$ if and only if
\begin{equation}\label{pppcondition1}
\sum\limits_{j=0}^{2N-1}B_jp_j^{(k)}(a_i-1)=0
\end{equation}
for $0\leq k\leq n_i-1$. Set
$$
\xi_k=\sum\limits_{j=0}^{2N-1}\varphi_jp_j^{(k)}(a_i-1).
$$
Condition (\ref{pppcondition1}) implies that
$\varphi\in(\Hc_N)_{a_i}$ if and only if $\varphi$ is orthogonal
to all $\xi_k$. Thus, $\xi_k$'s span $(\Hc_N)^{\perp}_{a_i}$.

Clearly, $(\Hc_N)_{\infty}$ is the span of $\varphi_k$ for $0\leq
k\leq 2N-n_{\infty}-1$, where $n_{\infty}$ is the order of
$\frac{w(x-1)}{w(x)}$ at $\infty$. Therefore
$(\Hc_N)_{\infty}^{\perp}$ is the span of $\varphi_k$ for $k\geq
2N-n_{\infty}$. We conclude that the dimension of the span of all
$(\Hc_N)^{\perp}_{a_i}$ (including $(\Hc_N)^{\perp}_{\infty}$) is
at most $n_{\infty}+\sum\limits_{i=1}^ln_{a_i}$. Pick an
orthonormal basis $\psi_1,\ldots ,\psi_n$ of this span. Note that
$n\leq n_{\infty}+\sum\limits_{i=1}^ln_{a_i}$.

 Since $\psi_1,\ldots ,\psi_n$ is a basis of $\mathcal{N}_D^{\perp}$, we
 have
$L_D=\sum\limits_{i=1}^nL_D\psi_i\otimes\psi_i$. Therefore we
obtain
$$
L_D=(I-K_N)D=\sum\limits_{i=1}^n(I-K_N)D\psi_i\otimes\psi_i.
$$
Set $\tilde{\psi}_i=(I-K_N)D\psi_i$. Since $L_D\psi_i\neq 0$ we
conclude that $\tilde{\psi}_i\in\Hc_N^{\perp}$, and we have
$$
[D,K_N]K_N=\sum\limits_{i=1}^n\tilde{\psi}_i\otimes\psi_i,\;\;\tilde{\psi}_i\in\Hc_N^{\perp},
\psi_i\in\Hc_N.
$$
Thus the first part of  Theorem \ref{THEOREM751} is proved.

 By
the antisymmetry of $\epsilon$ we obtain
$$
[D,K_N]K_N\epsilon=-\sum\limits_{i=1}^n\tilde{\psi}_i\otimes\epsilon\psi_i.
$$
In order to find $D_{\Hc_N}S_{N4}$ we need to compute the inverse
of $I-[D,K_N]K_N\epsilon$. Using the formula
$$
(I+\sum a_i\otimes
b_i)^{-1}=I-\sum\limits_{i,j}(T^{-1})_{ij}a_i\otimes b_j,
$$
where $T$ is the identity matrix plus the matrix of inner products
$(b_i,a_j)$ (assuming $T$ is invertible), and $\sum a_i\otimes
b_i$ is a finite rank operator, we obtain the formula for
$D_{\Hc_N}S_{N4}$ stated in Theorem \ref{THEOREM751}. \qed
\section{Proof of Theorem \ref{MainTheoremOO}.}
Set $L_{\epsilon}=(I-K_N)\epsilon $, and agree that the domain of
this operator is $\Hc_N$. Denote by $\mathcal{N}_{\epsilon}$ the
null space of $L_{\epsilon}$. Observe that
$$
\mathcal{N}_{\epsilon}=\{v\vert v\in\Hc_N, \epsilon v\in\Hc_N\}.
$$
Assume that $\psi\in\mathcal{N}_D^{\perp}$, and
$u\in\mathcal{N}_{\epsilon}$. Since $u\in\Hc_N$, and $K_N$ is a
symmetric operator, we have
$(u,K_N\epsilon\psi)=(u,\epsilon\psi)$. Since
$u\in\mathcal{N}_{\epsilon}$, $\epsilon u$ is an element of
$\Hc_N$ (see above). But $D\epsilon u=u\in\Hc_N$, which implies
that $\epsilon u$ belongs to $\mathcal{N}_D$, see previous
Section. Therefore, $(u,\epsilon\psi)=-(\epsilon u,\psi)=0$, and
$$
(u,K_N\epsilon\psi)=0,\;\hbox{for all
$\psi\in\mathcal{N}_D^{\perp}$ and $u\in\mathcal{N}_{\epsilon}$}.
$$
We conclude that $K_N\epsilon K_N$ takes $\mathcal{N}_D^{\perp}$
into $\mathcal{N}_\epsilon^{\perp}$.

Let us show that $\mathcal{N}_\epsilon^{\perp}$ has the same
dimension as $\mathcal{N}_D^{\perp}$. The operators $K_N\epsilon
K_N$ and $K_NDK_N$ are invertible because we have already shown
that their matrices $M^{(1)}_{jk}$ and $M_{jk}^{(4)}$ in the basis
$\varphi_0,\ldots,\varphi_{2N-1}$ are invertible, see Propositions
\ref{MP} and \ref{M1P}. We just saw that $K_N\epsilon K_N$ takes
$\mathcal{N}_D^{\perp}$ to $\mathcal{N}_{\epsilon}^{\perp}$, and
very similar arguments show that $K_ND K_N$ takes
$\mathcal{N}_\epsilon^{\perp}$ to $\mathcal{N}_{D}^{\perp}$.
Therefore
$\dim\mathcal{N}_D^{\perp}=\dim\mathcal{N}_{\epsilon}^{\perp}$.

Now take $\psi_1,\ldots,\psi_n$ found in the previous section and
orthonormalize the functions $K_N\epsilon\psi_j$. As a result
obtain an orthonormal basis in $\mathcal{N}_{\epsilon}^{\perp}$.
Denote this basis $\eta_1,\ldots,\eta_n$.

 This gives
$$
L_{\epsilon}=(I-K_N)\epsilon
=\sum\limits_{i=1}^n(I-K_N)\epsilon\eta_i\otimes\eta_i.
$$
Set $\tilde{\eta}_i=(I-K_N)\epsilon\eta_i$. By the same argument
as in the previous section  we find (assuming invertibility of U)
that
$$
\epsilon_{\Hc_N}S_{N1}=K_N-\sum\limits_{i,j=1}^n(U^{-1})_{ij}(\tilde{\eta}_i)\otimes\left(K_ND\eta_j\right),
$$
where the matrix $U$ is defined by its matrix elements by
$$
U_{ij}=\delta_{ij}+(D\eta_i,\tilde{\eta}_j).
$$
This immediately gives the formula in the statement of Theorem
\ref{MainTheoremOO}. \qed

\section{Proofs of Corollary \ref{COROLLARY1} and Corollary \ref{COROLLARY1ORTH}}
\begin{prop}\label{Propositionepsilonpsi1}
If the commutation relation between the operators $D$ and $K_N$
takes the form as in the statement of Corollary \ref{COROLLARY1},
with $\psi_1\in\Hc_N$, and $\psi_2\in\Hc_N^{\perp}$, then
$\epsilon\psi_1$ is an element  of $\Hc_N$.
\end{prop}
\begin{proof}
The explicit form of $[D,K_N]$ in the statement of Corollary
\ref{COROLLARY1} implies
\begin{equation}\label{KommutatorEpsilonKN}
[\epsilon,K_N]=\lambda(\epsilon\psi_1\otimes\epsilon\psi_2+\epsilon\psi_2\otimes\epsilon\psi_1).
\end{equation}
(We have used the fact that $\epsilon D=1$, and the antisymmetry
of $\epsilon$). Acting by $[\epsilon,K_N]$ on $\psi_1$ we find
$$
[\epsilon,K_N]\psi_1=\lambda\epsilon\psi_1(\epsilon\psi_2,\psi_1).
$$
This gives
$$
K_N\epsilon\psi_1=(1+\lambda(\psi_2,\epsilon\psi_1))\epsilon\psi_1.
$$
Therefore, either $\epsilon\psi_1$ is in $\Hc_N$, or
$K_N\epsilon\psi_1=0$. To rule out the second possibility observe
that $[D,K_N]K_N=\lambda\psi_2\otimes\psi_1$, as it follows from
the expression for $[D,K_N]$ in the hypothesis of Corollary
\ref{COROLLARY1}. Here $\psi_2\in\Hc_N^{\perp}$, and
$\psi_1\in\Hc_N$, and Theorem \ref{THEOREM751} implies that
$\psi_1$ is an element of $\mathcal{N}_D^{\perp}$. It was shown in
the proof of Theorem \ref{MainTheoremOO} that $K_N\epsilon$
bijectively maps $\mathcal{N}_D^{\perp}$ into
$\mathcal{N}_{\epsilon}^{\perp}$. Therefore, $K_N\epsilon\psi_1$
must be an element of $\mathcal{N}_{\epsilon}^{\perp}$, so
$K_N\epsilon\psi_1$ cannot be zero.
\end{proof}
\textbf{Proof of Corollary \ref{COROLLARY1}.} If the conditions of
Corollary \ref{COROLLARY1} are satisfied we have
$$
[D,K_N]K_N\epsilon=-\lambda\psi_2\otimes\epsilon\psi_1.
$$
Then Theorem \ref{MainTheorem} implies
$$
D_{\Hc_N}S_{N4}=(I_{\Hc_N+D\Hc_N}+\lambda\psi_2\otimes\epsilon\psi_1)^{-1}K_N.
$$
Now we use the formula
$$
(I+a\otimes b)^{-1}=I-T^{-1}a\otimes b,
$$
where $a=\lambda\psi_2$, $b=\epsilon\psi_1$, and
$T=1+\lambda(\psi_2,\epsilon\psi_1)$. By Proposition
\ref{Propositionepsilonpsi1} $\epsilon\psi_1\in\Hc_N$, and we
conclude that $T=1$. Therefore
$$
D_{\Hc_N}S_{N4}=K_N-\lambda\psi_2\otimes\epsilon\psi_1,
$$
or
$$
S_{N4}=\epsilon K_N-\lambda\epsilon\psi_2\otimes\epsilon\psi_1.
$$
\qed\\
\textbf{Proof of Corollary \ref{COROLLARY1ORTH}.}\\
By Theorem \ref{MainTheoremOO} we need to compute inverse of
$$
I-[\epsilon,K_N]K_ND.
$$
From equation (\ref{KommutatorEpsilonKN}) we find
$$
[\epsilon,K_N]K_ND=-\lambda(\epsilon\psi_1\otimes
DK_N\epsilon\psi_2+\epsilon\psi_2\otimes DK_N\epsilon\psi_1).
$$
Now we compute  $DK_N\epsilon\psi_1$ as follows
\begin{equation}
\begin{split}
DK_N\epsilon\psi_1&=[D,K_N]\epsilon\psi_1+K_ND\epsilon\psi_1=[D,K_N]\epsilon\psi_1+\psi_1\\
&=\lambda(\psi_1\otimes\psi_2+\psi_2\otimes\psi_1)\epsilon\psi_1+\psi_1
=\psi_1.
\end{split}
\nonumber
\end{equation}
(We have used $(\psi_1,\epsilon\psi_1)=0$, and the fact that by
Proposition \ref{Propositionepsilonpsi1}
$(\psi_2,\epsilon\psi_1)=0$). In a similar way we compute
$DK_N\epsilon\psi_2$
\begin{equation}
\begin{split}
DK_N\epsilon\psi_2&=[D,K_N]\epsilon\psi_2+K_ND\epsilon\psi_2=[D,K_N]\epsilon\psi_2\\
&=\lambda(\psi_1\otimes\psi_2+\psi_2\otimes\psi_1)\epsilon\psi_2
=0.
\end{split}
\nonumber
\end{equation}
Therefore we have obtained
$$
I-[\epsilon,K_N]K_ND=I+\lambda\epsilon\psi_2\otimes\psi_1.
$$
Using the formula
$$
(I+a\otimes b)^{-1}=I-T^{-1} a\otimes b
$$
where now $a=\lambda\epsilon\psi_2$, $b=\psi_1$,
$$
T=1+(b,a)=1+(\psi_1,\lambda\epsilon\psi_2)=1-\lambda(\epsilon\psi_1,\psi_2)=1,
$$
we obtain
$$
(I-[\epsilon,K_N]K_ND)^{-1}=I-\lambda\epsilon\psi_2\otimes\psi_1.
$$
This gives
$$
\epsilon_{\Hc_N}S_{N1}=K_N-\lambda\epsilon\psi_2\otimes\psi_1,
$$
and we finally arrive to the formula
$$
S_{N1}=DK_N-\lambda\psi_2\otimes\psi_1.
$$
\qed

\section{Discrete Riemann-Hilbert Problems (DRHP) and  difference
equations for orthogonal polynomials.}\label{SectionDRHP} In this
section we recall a few claims from Borodin and Boyarchenko
\cite{borodinboyarchenko}.
 Let $\X$
be a discrete locally finite subset of $\C$. Let $w:
\X\rightarrow\C$ be a function. Assume that all moments of $w$ are
finite. Denote by $\C[\zeta]$ the space of polynomials in the
complex variable $\zeta$, and introduce the inner product in
$\C[\zeta]$
$$
\left(f(\zeta),g(\zeta)\right)_w:=\sum\limits_{x\in\X}f(x)g(x)w(x).
$$
If the restriction of $(.,.)_w$ to the space $\C[\zeta]^{\leq d}$
of the polynomials of degree at most $d$ is non-degenerate for all
$d\geq 0$, then there exists a unique collection of monic
orthogonal polynomials $\{P_n(\zeta)\}_{n=0}^{\infty}$ associated
to $w$ such that if $m\neq n$, then $(P_n(\zeta),P_m(\zeta))_w=0$,
and $(P_n(\zeta),P_n(\zeta))_w\neq 0$ for all $n$. If this
condition holds, we say that the weight function $w$ is
non-degenerate. Note that if $\X$ is finite, and consists of $N+1$
points, the inner product $(.,.)_w$ is necessarily degenerate on
$\C[\zeta]^{\leq d}$ for all $d>N$. In this case we require that
$(.,.)_w$ be non-degenerate on $\C[\zeta]^{\leq d}$ for $0\leq
d\leq N$, and we are only interested in a collection of orthogonal
poynomials of degrees up to $N$.

As in Refs. \cite{borodin1,borodin2,borodinboyarchenko} , we say
that an analytic function
$$
m: \C\backslash\X\rightarrow Mat(2,\C)
$$
solves the DRHP $(\X,w)$ if $m$ has simple poles at the points of
$\X$ and its residues at these points are given by the jump (or
residue) condition
$$
\underset{\zeta=x}{\Res}\; m(\zeta)=\underset{\zeta\rightarrow
x}{\lim}\left(m(\zeta)\varpi(x)\right),\;\; x\in\X,
$$
where
$$
\varpi(x)=\left(
            \begin{array}{cc}
              0 & w(x) \\
              0 & 0 \\
            \end{array}
          \right).
$$
\begin{thm}
Let $\{P_n(\zeta)\}_{n=0}^N$ be  the collection of monic
orthogonal polynomials associated to $w$, where
$N=\mbox{card}(\X)-1\in\Zp\cup\left\{\infty\right\}$. For any
$k=1,2,\ldots, N$ the DRHP($\X,w$) has a unique solution
$m_{\X}(\zeta)$ satisfying the asymptotic condition
$$
m_{\X}(\zeta)\left(\begin{array}{cc}
                     \zeta^{-k} & 0 \\
                     0 & \zeta^k
                   \end{array}
\right)=I+O\left(\frac{1}{\zeta}\right),
$$
where $I$ is the identity matrix. If we write
$$
m_{\X}(\zeta)=\left(\begin{array}{cc}
                      m_{\X}^{11}(\zeta) & m_{\X}^{12}(\zeta) \\
                      m_{\X}^{21}(\zeta) & m_{\X}^{22}(\zeta)
                    \end{array}
\right),
$$
then $m_{\X}^{11}(\zeta)=P_k(\zeta)$,
$m_{\X}^{21}(\zeta)=(P_{k-1},P_{k-1})_{w}^{-1}P_{k-1}(\zeta)$,
$m_{\X}^{21}(\zeta)=\sum\limits_{x\in\X}\frac{P_k(\zeta)w(x)}{\zeta-x}$,
and
$m_{\X}^{22}(\zeta)=(P_{k-1},P_{k-1})_{w}^{-1}\sum\limits_{x\in\X}\frac{P_{k-1}(\zeta)w(x)}{\zeta-x}$.
\end{thm}
\begin{proof}
See Borodin and Boyarchenko \cite{borodinboyarchenko}, Section
2.3.
\end{proof}
Let $\X$ be a finite or a locally finite subset of $\R$,
$\mbox{card}(\X)=N+1$, where $N\in\Zp\cup\left\{\infty\right\}$.
Assume that $\X$ is parameterized as
$\X=\left\{\pi_x\right\}_{x=0}^N$, and that there exists an affine
transformation $\sigma:\C\rightarrow\C$ such that
$\sigma\pi_{x+1}=\pi_x$ for all $0\leq x< N$. Denote the
derivative of $\sigma$ (which is constant) by $\eta$, so that
$$
\sigma(\zeta_1)-\sigma(\zeta_2)=\eta(\zeta_1-\zeta_2)\;\;\mbox{for
all}\;\zeta_1,\zeta_2\in\C.
$$
Let $w$ be a strictly positive real valued function defined on
$\X$. Then $w$ is non-degenerate.
\begin{thm}\label{DFT}
Assume that there exist entire functions $d_1(\zeta)$ and
$d_2(\zeta)$ such that
$$
\frac{w(\pi_{x-1})}{w(\pi_x)}=\eta\cdot\frac{d_1(\pi_x)}{d_2(\pi_x)},\;\;1\leq
x\leq N,
$$
$$
d_1(\pi_0)=0,
$$
$$
d_2(\sigma^{-1}\pi_N)=0,\;\mbox{if $N$ is finite}.
$$
a) We have
$$
P_k(\sigma\zeta)=\left(M^{11}(\zeta)P_k(\zeta)+cM^{12}(\zeta)P_{k-1}(\zeta)\right)d_1(\zeta)^{-1}
$$
$$
cP_{k-1}(\sigma\zeta)=\left(M^{21}(\zeta)P_k(\zeta)+cM^{22}(\zeta)P_{k-1}(\zeta)\right)d_1^{-1}(\zeta)
$$
where $c=(P_{k-1},P_{k-1})_w^{-1}$, and $M^{11}(\zeta),
M^{12}(\zeta), M^{21}(\zeta), M^{22}(\zeta)$ are entire functions.
Moreover, the matrix $M(\zeta)$ whose entries are $M^{11}(\zeta)$,
$M^{12}(\zeta)$, $M^{21}(\zeta)$, $M^{22}(\zeta)$ is given by
$$
M(\zeta)=m_{\X}(\sigma\zeta)D(\zeta)m_{\X}^{-1}(\zeta),
$$
where
$$
D(\zeta)=\left(\begin{array}{cc}
           d_1(\zeta) & 0 \\
           0 & d_2(\zeta)
         \end{array}\right),
         $$
         and $m_{\X}(\zeta)$ is the solution of the DRHP($\X,w$)
         with the asymptotic condition
$$
m_{\X}(\zeta)\left(\begin{array}{cc}
                     \zeta^{-k} & 0 \\
                     0 & \zeta^k
                   \end{array}
\right)=I+O\left(\frac{1}{\zeta}\right).
$$
b) If it is known that $d_1(\zeta)$ and $d_2(\zeta)$ are
polynomials of degree at most $n$ in $\zeta$, and
$$
d_1(\zeta)=\lambda_1\zeta^n+(\mbox{lower terms}),
$$
$$
d_2(\zeta)=\lambda_2\zeta^n+(\mbox{lower terms}),
$$
then $2\times 2$ matrix $M(\zeta)$ is a polynomial of degree at
most $n$ in $\zeta$, and
$$
\left(\begin{array}{cc}
        M^{11}(\zeta) & M^{12}(\zeta) \\
        M^{21}(\zeta) & M^{22}(\zeta)
      \end{array}
\right)=\left(\begin{array}{cc}
                \eta^k\lambda_1 & 0 \\
                0 & \eta^{-k}\lambda_2
              \end{array}
\right)\zeta^k+(\mbox{lower terms}).
$$
\end{thm}
\begin{proof}
For the proof of these statements see Borodin and Boyarchenko
\cite{borodinboyarchenko}, Sections 3.1-3.3.
\end{proof}

\section{Commutation relations}
In this section set $\X=\Zp$, and $\sigma\zeta=\zeta-1$ for any
$\zeta\in\C$. Assume that  $w(x)$ is a strictly positive  function
on $\Zp$, which satisfies the condition in Theorem \ref{DFT}, i.e.
the  ratio of $w(x-1)$ and $w(x)$ equals the ratio of entire
functions $d_1(x)$ and $d_2(x)$, and $d_1(0)=0$.
\begin{prop}\label{35}
We have for $n=1,2,\ldots $
$$
\left(D_-\varphi_{n}\right)(x)=
\frac{M^{11}(x)}{d_2(x)}\varphi_{n}(x)+
\frac{M^{12}(x)}{d_2(x)(P_n,P_{n})_w^{1/2}(P_{n-1},P_{n-1})_w^{1/2}}\varphi_{n-1}(x),
$$
$$
\left(D_-\varphi_{n-1}\right)(x)=
\frac{M^{21}(x)(P_n,P_{n})_w^{1/2}(P_{n-1},P_{n-1})_w^{1/2}}{d_2(x)}\varphi_{n}(x)+
\frac{M^{22}(x)}{d_2(x)}\varphi_{n-1}(x).
$$
\end{prop}
\begin{proof}
These equations are equivalent to the difference equations for the
monic orthogonal polynomials in Theorem \ref{DFT}, a).
\end{proof}
\begin{prop}
Introduce the $2\times 2$ matrices $D_{\pm}(x)$ by the formula
\begin{equation}\label{D+-}
\left(\begin{array}{c}
        D_{\pm}\varphi_{2N}(x)  \\
        D_{\pm}\varphi_{2N-1}(x)
      \end{array}
\right)=\left(\begin{array}{cc}
                D^{11}_{\pm}(x) & D^{12}_{\pm}(x) \\
                D^{21}_{\pm}(x) & D^{22}_{\pm}(x)
              \end{array}
\right)\left(\begin{array}{c}
        \varphi_{2N}(x)  \\
        \varphi_{2N-1}(x)
      \end{array}
\right).
\end{equation}
The matrix elements of $D_+(x)$ are related  with the matrix
elements of $D_-(x+1)$ as
\begin{equation}\label{34DD}
\begin{split}
D^{11}_+(x)=D_-^{22}(x+1),\;\;D^{12}_+(x)=-D_-^{12}(x+1),\\
D^{21}_+(x)=-D_-^{21}(x+1),\;\;D^{22}_+(x)=D_-^{11}(x+1).
\end{split}
\end{equation}
\end{prop}
\begin{proof}
The definition of $D_{\pm}$ implies that
\begin{equation}\label{DDDDDD}
D_+(x)D_-(x+1)=\frac{w(x)}{w(x+1)}.
\end{equation}
Observe  that $\det M(x)=d_1(x)\cdot d_2(x)$ (see
\cite{borodinboyarchenko}, \S 3). This fact together with the
assumption that the  ratio of $w(x-1)$ and $w(x)$ equals the ratio
of  $d_1(x)$ and $d_2(x)$ imply that the determinant of the matrix
$D_-(x+1)$ equals $\frac{w(x)}{w(x+1)}$, which (together with
equation (\ref{DDDDDD})) leads to the relation between the matrix
elements in the statement of the Proposition.
\end{proof}
\begin{prop}\label{PropositionformulaDK}
We have
\begin{equation}
\begin{split}
&[D,K_N](x,y)\\&=a_{2N}\left(\varphi_{2N}(x),
\varphi_{2N-1}(x)\right)\left(
               \begin{array}{cc}
                 \frac{D^{21}_-(x+1)-D^{21}_-(y)}{x+1-y} & \frac{D^{22}_-(x+1)-D^{22}_-(y)}{x+1-y} \\
                 -\frac{D^{11}_-(x+1)-D^{11}_-(y)}{x+1-y} & -\frac{D^{12}_-(x+1)-D^{12}_-(y)}{x+1-y}\\
               \end{array}
             \right)
\left(\begin{array}{c}
               \varphi_{2N}(y) \\
               \varphi_{2N-1}(y)
             \end{array}
\right)\\
&+a_{2N}\left(\varphi_{2N}(x), \varphi_{2N-1}(x)\right)\left(
               \begin{array}{cc}
                 \frac{D^{21}_-(x)-D^{21}_-(y+1)}{x-y-1} & -\frac{D^{11}_-(x)-D^{11}_-(y+1)}{x-y-1} \\
                 \frac{D^{22}_-(x)-D^{22}_-(y+1)}{x-y-1} & -\frac{D^{12}_-(x)-D^{12}_-(y+1)}{x-y-1}\\
               \end{array}
             \right)
\left(\begin{array}{c}
               \varphi_{2N}(y) \\
               \varphi_{2N-1}(y)
             \end{array}
\right),
\end{split}
\nonumber
\end{equation}
where $a_{2N}$ is the coefficient in the Christoffel-Darboux
formula for the kernel $K(x,y)$,
$$
K(x,y)=a_{2N}\frac{\varphi_{2N}(x)\varphi_{2N-1}(y)-\varphi_{2N-1}\varphi_{2N}(y)}{x-y}.
$$
\end{prop}
\begin{proof}
A direct computation of the commutation relation between the
operators $D$ and $K$ gives
\begin{equation}
\begin{split}
[D,K](x,y)&=a_{2N}\biggl[\frac{\left(D_+\varphi_{2N}\right)(x)\varphi_{2N-1}(y)-
\left(D_+\varphi_{2N-1}\right)(x)\varphi_{2N}(y)}{x+1-y}\\
&-\frac{\left(D_-\varphi_{2N-1}\right)(y)\varphi_{2N}(x)
-\left(D_-\varphi_{2N}\right)(y)\varphi_{2N-1}(x)}{x+1-y}\\
&-\frac{\left(D_-\varphi_{2N}\right)(x)\varphi_{2N-1}(y)-
\left(D_-\varphi_{2N-1}\right)(x)\varphi_{2N}(y)}{x-1-y}\\
&+\frac{\left(D_+\varphi_{2N-1}\right)(y)\varphi_{2N}(x)-
\left(D_+\varphi_{2N}\right)(y)\varphi_{2N-1}(x)}{x-1-y}\biggr]
\end{split}
\nonumber
\end{equation}
Express $D_{\pm}\varphi_{2N}, D_{\pm}\varphi_{2N-1}$ as linear
combinations of $\varphi_{2N}$ and $\varphi_{2N-1}$ whose
coefficients are entries of matrices $D_+(x), D_-(x)$. In the
resulting formula replace the matrix elements $D^{11}_+(x)$,
$D^{12}_+(x)$, $D^{21}_+(x)$, $D^{22}_+(x)$ by the matrix elements
$D^{11}_-(x+1)$, $D^{12}_-(x+1)$, $D^{21}_-(x+1)$, $D^{22}_-(x+1)$
in accordance with formula (\ref{34DD}). This gives the formula in
the statement of the Proposition.
\end{proof}
\begin{prop}\label{PROPOSITION569568}
Assume that the functions $d_1$, $d_2$ are polynomials of degree
at most $m$. Furthermore, assume that $d_2$ has zeros at points
$a_1,\ldots, a_l$ of degrees $n_1,\ldots,n_l$ correspondingly.
Then each of the expressions
$\frac{D_-^{21}(x+1)-D^{21}_-(y)}{x+1-y}$,
$\frac{D_-^{22}(x+1)-D^{22}_-(y)}{x+1-y}$,
$\frac{D_-^{11}(x+1)-D^{11}_-(y)}{x+1-y}$,
$\frac{D_-^{12}(x+1)-D^{21}_-(y)}{x+1-y}$ is a finite sum of
expressions of the form
$$
\left(\sum\limits_{k=0}^{m-1-n_1-\ldots-n_l}A_kx^k+\sum\limits_{i=1}^l\sum\limits_{k_i=1}^{n_i}
\frac{B_{k_i}}{(x+1-a_i)^{k_i}}\right)
\left(\sum\limits_{k=0}^{m-1-n_1-\ldots-n_l}C_ky^k+\sum\limits_{i=1}^l\sum\limits_{k_i=1}^{n_i}
\frac{D_{k_i}}{(y-a_i)^{k_i}}\right),
$$
and each of the expressions
$\frac{D_-^{21}(x)-D^{21}_-(y+1)}{x-y-1}$,
$\frac{D_-^{11}(x)-D^{11}_-(y+1)}{x-y-1}$,
$\frac{D_-^{22}(x)-D^{22}_-(y+1)}{x-y-1}$,
$\frac{D_-^{12}(x)-D^{12}_-(y+1)}{x-y-1}$ is a finite sum of
expressions of the form
$$
\left(\sum\limits_{k=0}^{m-1-n_1-\ldots-n_l}\tilde{A}_kx^k+\sum\limits_{i=1}^l\sum\limits_{k_i=1}^{n_i}
\frac{\tilde{B}_{k_i}}{(x-a_i)^{k_i}}\right)
\left(\sum\limits_{k=0}^{m-1-n_1-\ldots-n_l}\tilde{C}_ky^k+\sum\limits_{i=1}^l\sum\limits_{k_i=1}^{n_i}
\frac{\tilde{D}_{k_i}}{(y+1-a_i)^{k_i}}\right),
$$
where $A_k$, $C_k$, $B_{k_i}$ and $D_{k_i}$, $\tilde{A}_k$,
$\tilde{C}_k$, $\tilde{B}_{k_i}$ and $\tilde{D}_{k_i}$ are some
constant coefficients.
\end{prop}
\begin{proof}
If $d_1(x)$ and $d_2(x)$ are polynomials of degree at most $m$,
then $M^{11}(x)$, $M^{12}(x)$, $M^{21}(x)$, and $M^{22}(x)$ are
polynomials of degree at most $m$, see Theorem \ref{DFT}, b). By
Proposition \ref{35}, and by the very definition of the matrix
$D_-(x)$ each $D_-^{11}(x)$, $D_-^{22}(x)$, $D_-^{12}(x)$,
$D_-^{21}(x)$ is a polynomial of degree at most $m$ divided by
$d_2(x)$. Set
$$
d_2(x)=\const\cdot(x-a_1)^{n_1}\ldots (x-a_l)^{n_l},
$$
where
$$
n_1+n_2+\ldots+n_l\le m.
$$
Denote $D^{11}_-(x)d_2(x)=A_mx^m+\ldots +A_0$. Then we have
\begin{equation}
\begin{split}
\frac{D^{11}_-(x+1)-D^{11}_-(y)}{x+1-y}&=\frac{A_m(x+1)^m+\ldots+A_1(x+1)+A_0}{(x+1-y)(x+1-a_1)^{n_1}\ldots
(x+1-a_l)^{n_l}}\\
&-\frac{A_my^m+\ldots+A_1y+A_0}{(x+1-y)(y-a_1)^{n_1}\ldots
(y-a_l)^{n_l}}.
\end{split}
\nonumber
\end{equation}
Since
$$
\frac{(x+1)^k-y^k}{x+1-y}=\sum\limits_{a+b\leq k-1}(x+1)^ay^b
$$
we arrive at the result. The same considerations are applicable
to all expressions in the statement of the Proposition.
\end{proof}

\section{Difference equations for the orthonormal functions
associated to the Meixner and to the Charlier weights}
\begin{prop}\label{SYSTEMMEIXNER1}
Set
$$
\varphi_n(x)=\left(P_n,P_n\right)^{-1}_{w_{Meixner}}P_n(x)\sqrt{w_{Meixner}(x)}
$$
where $w_{Meixner}(x)$ is the Meixner weight defined by equation
(\ref{MeinerWeight171}), and $\{P_j(x)\}$ is the family of the
monic Meixner polynomials. The functions
$\left\{\varphi_n(x)\right\}_{n=0}^{\infty}$ satisfy the following
system of difference equations
\begin{equation}
  \begin{array}{ll}
    \sqrt{cx(x+\beta-1)}\varphi_n(x-1)=(x-n)\varphi_n(x)+\sqrt{cn(n+\beta-1)}\varphi_{n-1}(x), \\
    \sqrt{cx(x+\beta-1)}\varphi_{n-1}(x-1)=c(x+(n+\beta-1))\varphi_{n-1}(x)-\sqrt{cn(n+\beta-1)}\varphi_{n}(x). \\
  \end{array}
  \nonumber
\end{equation}
\end{prop}
\begin{proof}
Let $m(\zeta)$ denote the solution of DRHP($\Zp,w_{Meixner}$)
satisfying the asymptotic condition
$$
m(\zeta)\left(\begin{array}{cc}
                \zeta^{-n} & 0 \\
                0 & \zeta^{n}
              \end{array}
\right)=I+O({\zeta}^{-1}),
$$
see Section \ref{SectionDRHP}. The matrix $m(\zeta)$ has a full
asymptotic expansion in $\zeta$ as $\zeta\rightarrow\infty$, and
we can write
\begin{equation}\label{m0}
m(\zeta)\left(\begin{array}{cc}
                     \zeta^{-n} & 0 \\
                     0 & \zeta^{n}
                   \end{array}
\right)=I+\left(\begin{array}{cc}
                  \alpha & \tau \\
                  \gamma & \delta
                \end{array}
\right)\zeta^{-1}+O(\zeta^{-2}).
\end{equation}
By Theorem \ref{DFT} there exists an entire matrix-valued function
$M(\zeta)$  such that
\begin{equation}\label{m1}
M(\zeta)=m(\zeta-1) D(\zeta) m^{-1}(\zeta),
\end{equation}
where the matrix $D(\zeta)$ has the following form
$$
D(\zeta)=\left(\begin{array}{cc}
                 \zeta & 0 \\
                 0 & c\zeta+c(\beta-1)
               \end{array}
 \right).
$$
We can rewrite equation (\ref{m1}) as
\begin{equation}
\begin{split}
M(\zeta)&=\left[m(\zeta-1)\left(\begin{array}{cc}
                                 (\zeta-1)^{-n} & 0 \\
                                 0 & (\zeta-1)^n
                               \end{array}
\right)\right]\\
&\times \left[\left(\begin{array}{cc}
                      (1-\frac{1}{\zeta})^n & 0 \\
                      0 & (1-\frac{1}{\zeta})^{-n}
                    \end{array}
\right)\left(\begin{array}{cc}
               \zeta & 0 \\
               0 & c\zeta+c(\beta-1)
             \end{array}
\right)\right]\left[m(\zeta)\left(\begin{array}{cc}
                                 \zeta^{-n} & 0 \\
                                 0 & \zeta^n
                               \end{array}
\right)\right]^{-1}.
\end{split}
\nonumber
\end{equation}
The last multiplier in the expression above has the form
$$
\left(
\begin{array}{cc}
  1-\alpha\zeta^{-1} & -\tau\zeta^{-1} \\
  -\gamma\zeta^{-1} & 1-\delta\zeta^{-1}
\end{array}
\right)+O(\zeta^{-2}),
$$
the multiplication of the two matrices in the middle  gives
$$
\left(\begin{array}{cc}
        \zeta-n & 0 \\
        0 & c\zeta+cn+c(\beta-1)
      \end{array}
\right)+O(\zeta^{-1}),
$$
and the result of the multiplication of the first two matrices in
the expression for $M(\zeta)$ is
$$
\left(\begin{array}{cc}
        1+\alpha\zeta^{-1} & \tau\zeta^{-1} \\
        \gamma\zeta^{-1} & 1+\delta\zeta^{-1}
      \end{array}
\right)+O(\zeta^{-2}),
$$
as it is evident from (\ref{m0}). The straightforward calculation
gives
$$
M(\zeta)=\left(\begin{array}{cc}
                 \zeta-n & (c-1)\tau \\
                 -(c-1)\gamma & c\zeta+c(n+\beta-1)
               \end{array}
\right)+O(\zeta^{-1}).
$$
Since $M(\zeta)$ is entire the last term $O(\zeta^{-1})$ is
identically zero by Liouville's theorem. By Theorem \ref{DFT} the
monic Meixner polynomials satisfy the following system of the
difference equations
\begin{equation}\label{m3}
\begin{split}
&\zeta P_n(\zeta-1)=(\zeta-n)P_n(\zeta)+c_{n-1}(c-1)\tau
P_{n-1}(\zeta)\\
&c_{n-1}\zeta P_{n-1}(\zeta-1)=-(c-1)\gamma
P_n(\zeta)+c_{n-1}\left(c\zeta+c(n+\beta-1)\right)P_{n-1}(\zeta)
\end{split}
\end{equation}
where $c_n=\left(P_n,P_n\right)_w^{-1}$. Moreover, we also have
the condition $\det M(\zeta)=\det D(\zeta)$ which follows from the
fact that the determinant of the solution of the Discrete
Riemann-Hilbert problem identically equals 1, and from the
relation between $M(\zeta)$ and $D(\zeta)$, equation (\ref{m1}).
This condition implies a relation between parameters,
\begin{equation}\label{m4}
(c-1)^2\tau\gamma=cn(n+\beta-1).
\end{equation}
Replacing $n-1$ by $n$ in the second equation of system
(\ref{m3}), and inserting the result into the first equation we
obtain the recurrence relation for the monic Meixner polynomials.
Since the recurrence equation for the Meixner polynomials is known
(see, for example, equation (1.9.3) in Ref. \cite{koekoek}) this
(together with formula (\ref{m4})) determines coefficients
$\gamma$ and $\tau$ in system (\ref{m3}). Finally, rewriting
(\ref{m3}) in terms of functions $\varphi_n(x)$ and
$\varphi_{n-1}(x)$ we obtain the difference equations in the
statement of the Proposition.
\end{proof}
\begin{prop}\label{PSystemCh}
Let $\left\{\varphi_n(x)\right\}_{n=0}^{\infty}$ be the family of
the orthonormal functions associated with the Charlier weight
defined by equation \ref{ChW172}.  The functions
$\left\{\varphi_n(x)\right\}_{n=0}^{\infty}$ satisfy the following
system of difference equations
\begin{equation}
\begin{split}
&\sqrt{ax}\varphi_n(x-1)=(x-n)\varphi_n(x)+\sqrt{an}\varphi_{n-1}(x),\\
&\sqrt{ax}\varphi_{n-1}(x-1)=-\sqrt{an}\varphi_{n}(x)+a\varphi_{n-1}(x).
\end{split}
\nonumber
\end{equation}
\end{prop}
\begin{proof}
The system of difference equations for the orthonormal functions
$\{\varphi_n(x)\}_{n=0}^{\infty}$ associated to the Charlier
weight follows from the corresponding system of difference
equations for the case of the Meixner weight. Indeed, we have the
following limiting relation
\begin{equation}\label{MeixnerCHlimitrelation}
\underset{n\rightarrow\infty}{\lim}M_n\left(x;\beta,\frac{a}{\beta+a}\right)=C_n(x;a)
\end{equation}
between the $n$th Charlier polynomial $C_n(x;a)$ and the $n$th
Meixner polynomial $M_n(x;\beta,c)$, see, for example, Ref.
\cite{ismail}.
\end{proof}

\section{The Meixner and Charlier symplectic and orthogonal ensembles}
By Propositions \ref{PropositionformulaDK},
\ref{PROPOSITION569568} the kernel of the operator $[D,K_N]$ is
expressible in terms of the following functions:
\begin{equation}\label{function0}
x^k\varphi_{2N-1}(x),\;\; x^k\varphi_{2N};\;\;\; 0\leq k\leq
m-1-\sum\limits_{i=1}^ln_i,
\end{equation}
and, for each zero $a_i$ of $d_2(x)$, the functions
\begin{equation}\label{function1}
(x+1-a_i)^{-k_i}\varphi_{2N-1}(x),\;\;(x+1-a_i)^{-k_i}\varphi_{2N}(x);\;\;
1\leq k_i\leq n_i,
\end{equation}
\begin{equation}\label{function2}
(x-a_i)^{-k_i}\varphi_{2N-1}(x),\;\;
(x-a_i)^{-k_i}\varphi_{2N}(x);\;\; 1\leq k_i\leq n_i.
\end{equation}
\begin{prop}
In the case of the Meixner weight we have
\begin{equation}\label{F}
\begin{split}
&[D,K_N](x,y)=\frac{\sqrt{2N(2N+\beta-1)}}{c-1}\left(\zeta_{2N}(x),
\zeta_{2N-1}(x),\eta_{2N}(x),\eta_{2N-1}(x)\right)\\
&\left(\begin{smallmatrix}
   0 & 0 & \sqrt{2N(2N+\beta-1)} & -2N\sqrt{c} \\
   0 & 0 & -\frac{2N-1+\beta}{\sqrt{c}} & \sqrt{2N(2N+\beta-1)} \\
   \sqrt{2N(2N+\beta-1)} & -\frac{2N-1+\beta}{\sqrt{c}} & 0 & 0 \\
   -2N\sqrt{c} & \sqrt{2N(2N+\beta-1)} & 0 & 0
 \end{smallmatrix}\right)
\left(
\begin{array}{c}
  \zeta_{2N}(y) \\
  \zeta_{2N-1}(y) \\
  \eta_{2N}(y) \\
  \eta_{2N-1}(y)
\end{array}\right)
\end{split}
\end{equation}
where we have introduced $\zeta_j(x)=\frac{\varphi_j(x)}{x+\beta}$
and $\eta_j(x)=\frac{\varphi_j}{x+\beta-1}$.
\end{prop}
\begin{proof}
We use Proposition \ref{PropositionformulaDK} which determines the
commutation relation between the operators $D$ and $K_N$ in terms of
the matrix elements of the matrix $D_-(x)$ defined by equation
(\ref{D+-}). This matrix can be written explicitly in the case of
the Meixner weight since we have obtained in Proposition
\ref{SYSTEMMEIXNER1} the system of difference equations  for the
orthonormal functions $\varphi_{2N-1}(x)$ and $\varphi_{2N}(x)$.
Namely,
$$
D_-(x)=\left[\begin{array}{cc}
               \frac{x-2N}{c(x+\beta-1)} & \frac{\sqrt{2N(2N+\beta-1)}}{\sqrt{c}(x+\beta-1)} \\
               -\frac{\sqrt{2N(2N+\beta-1)}}{\sqrt{c}(x+\beta-1)} &
\frac{x+(2N+\beta-1)}{x+\beta-1}
             \end{array}
\right]
$$
Inserting the matrix elements of the matrix $D_-(x)$ into the
formula in the statement of Proposition
\ref{PropositionformulaDK}, and taking into account that the
coefficient $a_{2N}$ corresponding to the case of the Meixner
weight equals $-\frac{\sqrt{2Nc(\beta+2N-1)}}{1-c}$ we obtain the
formula for $[D,K_N]$.
\end{proof}

We want to construct linear combinations
$\psi_1,\psi_2,\psi_3,\psi_4$ of
$\zeta_{2N},\zeta_{2N-1},\eta_{2N},\eta_{2N-1}$ in such a way that
$\psi_1$ and $\psi_2$ will be lying in $\mathcal{H}_N$, and
$\psi_3,\psi_4$ will be lying in $\mathcal{H}^{\perp}_N$.
\begin{prop}
Set
\begin{small}
\begin{equation}\label{P1}
\psi_1=\sqrt{2cN}\zeta_{2N}-\sqrt{\beta+2N-1}\zeta_{2N-1},
\end{equation}
\begin{equation}\label{P2}
\psi_2=\sqrt{2N}F(-2N+1,\beta-1;\beta;c)\eta_{2N}-\sqrt{c(\beta+2N-1)}F(-2N,\beta-1;\beta;c)\eta_{2N-1},
\end{equation}
\begin{equation}\label{P3}
\psi_3=(\beta+2N)
F(\beta,\beta+2N-1;\beta+2N;c)\zeta_{2N}-\sqrt{2cN(\beta+2N-1)}F(\beta,\beta+2N;\beta+2N+1;c)\zeta_{2N-1},
\end{equation}
\begin{equation}\label{P4}
\psi_4=\sqrt{\beta+2N-1}\eta_{2N}-\sqrt{2cN}\eta_{2N-1}.
\end{equation}
\end{small}
Then $\psi_1,\psi_2,\psi_3,\psi_4$ are linear independent,
$\psi_1, \psi_2$ are elements  of $\mathcal{H}_N$, and
$\psi_3,\psi_4$ are elements of $H^{\perp}_N$.
\end{prop}
\begin{proof}
a) Suppose that the linear combination
$$
\psi_1=A\frac{\varphi_{2N}(x)}{x+\beta}+B\frac{\varphi_{2N-1}(x)}{x+\beta}
$$
is an element of $\mathcal{H}_N$. Here $A$, $B$ are some
coefficients. If $\psi_1$ is lying in $\mathcal{H}_N$, then
\begin{equation}\label{pf1}
Ap_{2N}(-\beta)+Bp_{2N-1}(-\beta)=0
\end{equation}
Therefore, we need to check that the equality
\begin{equation}\label{pf2}
\frac{p_{2N-1}(-\beta)}{p_{2N}(-\beta)}=\sqrt{\frac{2cN}{\beta+2N-1}}
\end{equation}
is satisfied in order to conclude that $\psi_1$ defined by
(\ref{P1}) is lying in $\mathcal{H}_N$. To check (\ref{pf2})  note
that
$$
p_{n}(x)=\frac{M_n(x;\beta,c)}{\|M_n(.;\beta,c)\|},
$$
where $M_n(x;\beta,c)$ is the $n^{\mbox{th}}$ Meixner polynomial.
We have $M_n(x;\beta,c)=F(-n,-x;\beta;\frac{c-1}{c})$, see, for
example, Refs. \cite{ismail, koekoek} for the basic properties of
the Meixner polynomials. The norm, $\|M_n(.;\beta,c)\|$, is equal
to
$$\|M_n(.;\beta,c)\|=\sqrt{\frac{\Gamma(\beta)\Gamma(n+1)}{\Gamma(\beta+n)}} c^{-n/2}(1-c)^{-\beta/2}.$$
Taking into account the formula
$$
F(-n,\beta;\beta;-z)=(1+z)^n,
$$
we see that  relation (\ref{pf2}) indeed holds.

b) In the same way we show that $\psi_2$ defined by equation
(\ref{P2}) is an element of $\mathcal{H}_N$.

c) Now we need to prove that the functions $\psi_3, \psi_4$
defined by equations (\ref{P3}), (\ref{P4})
 are elements of $\mathcal{H^{\perp}}$. To show this we note that
the space $\mathcal{H}_N$ in the case of the Meixner weight can be
understood as that spanned by the functions
$$
\varphi_0;\;\varphi_1;\;
(x+\beta)(x+\beta-1)x^k\sqrt{w_{Meixner}(x)},\;\; 0\leq k\leq
2N-3.
$$
All linear combinations of $\frac{\varphi_{2N-1}(x)}{x+\beta}$,
$\frac{\varphi_{2N}(x)}{x+\beta}$,
$\frac{\varphi_{2N-1}(x)}{x+\beta-1}$ and
$\frac{\varphi_{2N}(x)}{x+\beta-1}$ are certainly orthogonal to
$(x+\beta)(x+\beta-1)x^k\sqrt{w_{Meixner}(x)},\; 0\leq k\leq
2N-3.$ Therefore, we can take as $\psi_3, \psi_4$ linear
combinations of the form
$$
C\frac{\varphi_{2N-1}(x)}{x+\beta}+D\frac{\varphi_{2N}(x)}{x+\beta}
+E\frac{\varphi_{2N-1}(x)}{x+\beta-1}+F\frac{\varphi_{2N}(x)}{x+\beta-1},
$$
where the coefficients $C, D, E$ and $F$ are subjected  to the
conditions
\begin{equation}\label{pf3}
C\left(\varphi_0,\frac{\varphi_{2N-1}(x)}{x+\beta}\right)+D\left(\varphi_0,\frac{\varphi_{2N}(x)}{x+\beta}\right)
+E\left(\varphi_0,\frac{\varphi_{2N-1}(x)}{x+\beta-1}\right)+F\left(\varphi_0,\frac{\varphi_{2N}(x)}{x+\beta-1}\right)=0,
\end{equation}
\begin{equation}\label{pf4}
C\left(\varphi_1,\frac{\varphi_{2N-1}(x)}{x+\beta}\right)+D\left(\varphi_1,\frac{\varphi_{2N}(x)}{x+\beta}\right)
+E\left(\varphi_1,\frac{\varphi_{2N-1}(x)}{x+\beta-1}\right)+F\left(\varphi_1,\frac{\varphi_{2N}(x)}{x+\beta-1}\right)=0.
\end{equation}
In particular, we can take
$$
\psi_3=C\;\frac{\varphi_{2N-1}}{x+\beta}+D\;\frac{\varphi_{2N}}{x+\beta},\;\;
\psi_4=E\;\frac{\varphi_{2N-1}}{x+\beta-1}+F\;\frac{\varphi_{2N}}{x+\beta-1},
$$
provided that the following conditions on the coefficients $C,D,E$
and $F$ are satisfied
\begin{equation}\label{pf5}
\frac{C}{D}= -\frac{\left(\varphi_0,
\frac{\varphi_{2N}}{x+\beta}\right)}{\left(\varphi_0,\frac{\varphi_{2N-1}}{x+\beta}\right)},\;\;
\frac{E}{F}= -\frac{\left(\varphi_0,
\frac{\varphi_{2N}}{x+\beta-1}\right)}{\left(\varphi_0,\frac{\varphi_{2N-1}}{x+\beta-1}\right)}.
\end{equation}
(It is not hard to check that for $n\geq 1$
\begin{equation}
\left(\varphi_1,\frac{\varphi_n}{x+\beta}\right)=
\sqrt{\frac{\beta}{c}}\left(\varphi_0,\frac{\varphi_n}{x+\beta}\right),
\;\;\left(\varphi_1,\frac{\varphi_n}{x+\beta-1}\right)=\frac{c+\beta-1}{\sqrt{\beta
c}} \left(\varphi_0,\frac{\varphi_n}{x+\beta-1}\right).
\end{equation}
Taking into account these relations  we see that if conditions
(\ref{pf5}) are satisfied, equations (\ref{pf3}), (\ref{pf4}) hold
as well).

Thus it remains to check that for $n\geq 1$ the following
relations are valid:
\begin{equation}\label{pf7}
\frac{\left(\varphi_0,
\frac{\varphi_{n}}{x+\beta}\right)}{\left(\varphi_0,\frac{\varphi_{n-1}}{x+\beta}\right)}=
\frac{\sqrt{cn(\beta+n-1)}F(\beta,\beta+n;\beta+n+1;c)}{(\beta+n)F(\beta,\beta+n-1;\beta+n;c)},
\end{equation}
\begin{equation}\label{pf8}
\frac{\left(\varphi_0,
\frac{\varphi_{n}}{x+\beta-1}\right)}{\left(\varphi_0,\frac{\varphi_{n-1}}{x+\beta-1}\right)}=
\frac{\sqrt{cn}}{(\beta+n-1)}.
\end{equation}
Let us compute the scalar product in the left-hand side of
equation (\ref{pf7}). We have
$$
\left(\varphi_0,\frac{\varphi_n}{x+\beta}\right)=(1-c)^{\beta}\sqrt{\frac{(\beta)_nc^n}{n!}}
\sum\limits_{x=0}^{+\infty}\frac{M_n(x;\beta,c)}{x+\beta}\;\frac{(\beta)_xc^x}{x!}.
$$
It is convenient to exploit the discrete  Rodrigues formula  for
the Meixner polynomials (see, for example, Ismail \cite{ismail},
Section 6.1):
$$
M_n(x;\beta,c)\frac{(\beta)_xc^x}{x!}=\nabla^{n}\left[\frac{(\beta+n)_xc^x}{x!}\right],
$$
where
$$
\left(\nabla f\right)(x)=f(x)-f(x-1).
$$
This gives
\begin{equation}\label{pf6}
\begin{split}
\left(\varphi_0,\frac{\varphi_n}{x+\beta}\right)=(1-c)^{\beta}\sqrt{\frac{(\beta)_nc^n}{n!}}
\sum\limits_{x=0}^{+\infty}\frac{1}{x+\beta}\sum\limits_{k=0}^n\left(\begin{array}{c}
                                                                        n \\
                                                                        k
                                                                      \end{array}
\right)(-1)^k\frac{(\beta+n)_{x-k}c^{x-k}}{(x-k)!},
\end{split}
\end{equation}
where we have used the formula
$$
\left(\nabla^{n}f\right)(x)=\sum\limits_{k=0}^n\left(\begin{array}{c}
                                                       n \\
                                                       k
                                                     \end{array}
\right)(-1)^kf(x-k).
$$
Two sums in the righthand side of equation (\ref{pf6}) can be
rewritten further as
$$
\sum\limits_{k=0}^n\left(\begin{array}{c}
                                                                        n \\
                                                                        k
                                                                      \end{array}
\right)\sum\limits_{x=0}^{+\infty}\frac{1}{x+\beta+k}\frac{(\beta+n)_xc^x}{x!}.
$$
Taking into account the formula
$$
\sum\limits_{k=0}^n(-1)^k\left(\begin{array}{c}
                                                                        n \\
                                                                        k
                                                                      \end{array}
\right) \frac{1}{x+k}=\frac{n!}{x(x+1)\ldots (x+n)}
$$
we sum up over $k$ and obtain
\begin{equation}
\begin{split}
\left(\varphi_0,\frac{\varphi_n}{x+\beta}\right)=&(1-c)^{\beta}\sqrt{\frac{(\beta)_nc^n}{n!}}
\sum\limits_{x=0}^{+\infty}\frac{n!}{(x+\beta)(x+\beta+1)\ldots(x+\beta+n)}\frac{(\beta+n)_xc^x}{x!}\\
&=\frac{(1-c)^{\beta}}{\beta+n}c^{n/2}\sqrt{\frac{\Gamma(\beta)\Gamma(n+1)}{\Gamma(\beta+n)}}
F(\beta,\beta+n;\beta+n+1;c).
\end{split}
\nonumber
\end{equation}
This formula implies that relation (\ref{pf7}) indeed holds.
 Relation (\ref{pf8}) can be checked in the same way.

\end{proof}

\begin{prop}\label{PPPP}
In the case of the Meixner weight the commutation relation between
$D$ and $K_N$ can be expressed as
\begin{equation}\label{pf10}
\left[D,K_N\right]=\frac{\sqrt{2N(2N+\beta-1)}}{(c-1)\sqrt{c}}\left(\psi_1,\psi_4\right)
\left(\begin{array}{cc}
        0 & 1 \\
        1 & 0
      \end{array}
\right)\left(\begin{array}{c}
               \psi_1 \\
               \psi_4
             \end{array}
\right),
\end{equation}
where $\psi_1\in\mathcal{H}_N$ and
$\psi_4\in\mathcal{H}^{\perp}_N$ are defined by equations
(\ref{P1}), (\ref{P4}) correspondingly.
\end{prop}
\begin{proof}
Equations (\ref{P1})-(\ref{P4}) can be rewritten as
\begin{equation}\label{pf11}
\left(\begin{array}{c}
        \psi_1 \\
        \psi_2 \\
        \psi_3 \\
        \psi_4
      \end{array}
\right)=\left(\begin{array}{cccc}
                1 & 0 & 0 & 0 \\
                0 & 0 & 1 & 0 \\
                0 & 1 & 0 & 0 \\
                0 & 0 & 0 & 1
              \end{array}
\right)\left(\begin{array}{cccc}
               m_{11} & m_{12} & 0 & 0 \\
               m_{21} & m_{22} & 0 & 0 \\
               0 & 0 & m_{33} & m_{34} \\
               0 & 0 & m_{43} & m_{44}
             \end{array}
\right)\left(\begin{array}{c}
               \zeta_{2N} \\
               \zeta_{2N-1} \\
               \eta_{2N} \\
               \eta_{2N-1}
             \end{array}
\right),
\end{equation}
where
$$m_{11}=\sqrt{2cN},\;\; m_{12}=-\sqrt{\beta+2N-1};
$$
$$
m_{21}=(\beta+2N)F(\beta,\beta+2N-1;\beta+2N;c);
$$
$$m_{22}=-\sqrt{2cN(\beta+2N-1)}F(\beta,\beta+2N;\beta+2N+1;c);
$$
$$
m_{33}=\sqrt{2N}F(-2N+1,\beta-1;\beta;c),\;\;
m_{34}=-\sqrt{c(\beta+2N-1)}F(-2N,\beta-1;\beta;c);
$$
$$
m_{43}=\sqrt{\beta+2N-1},\;\;m_{44}=-\sqrt{2cN}.
$$
Introduce the following notation
$$
M_1=\left(\begin{array}{cc}
            m_{11} & m_{12} \\
            m_{21} & m_{22}
          \end{array}
\right),\;\; M_2=\left(\begin{array}{cc}
                         m_{33} & m_{34} \\
                         m_{43} & m_{44}
                       \end{array}
\right),
$$
$$
B=\frac{\sqrt{2N(2N+\beta-1)}}{c-1}\left[\begin{array}{cc}
                                          b_{11} & b_{12} \\
                                          b_{21} & b_{22}
                                        \end{array}
\right],
$$
where $b_{11}=\sqrt{2N(2N+\beta-1)}$, $b_{12}=-2N\sqrt{c}$,
$b_{21}=-\frac{2N-1+\beta}{\sqrt{c}}$,
$b_{22}=\sqrt{2N(2N+\beta-1)}$, and
$$
Q=\left(\begin{array}{cccc}
          1 & 0 & 0 & 0 \\
          0 & 0 & 1 & 0 \\
          0 & 1 & 0 & 0 \\
          0 & 0 & 0 & 1
        \end{array}
\right).
$$
Using equation (\ref{pf11}) we express $\zeta_{2N}, \zeta_{2N-1},
\eta_{2N}$ and $\eta_{2N-1}$ in terms of $\psi_1,\psi_2,\psi_3$
and $\psi_4$, and rewrite formula (\ref{F}) as follows
$$
\left[D,K_N\right]=\left(\psi_1,\psi_2,\psi_3,\psi_4\right)Q\left(\begin{array}{cc}
                                                                  0 & (M_1^T)^{-1}BM_2^{-1} \\
                                                                  (M_2^T)^{-1}B^TM_1^{-1} & 0
                                                                \end{array}
\right)Q\left(\begin{array}{c}
                \psi_1 \\
                \psi_2 \\
                \psi_3 \\
                \psi_4
              \end{array}
\right).
$$
Now we compute the matrix $(M_1^T)^{-1}BM_2^{-1}$ explicitly. We
have
$$
(M_1^T)^{-1}BM_2^{-1}=\frac{\sqrt{2N(2N+\beta-1)}}{(c-1)\triangle_1\triangle_2}
\left(\begin{array}{cc}
  m_{22} & -m_{21} \\
  -m_{12} & m_{11} \\
\end{array}\right)
\left(\begin{array}{cc}
  b_{11} & b_{12} \\
  b_{21} & b_{22} \\
\end{array}\right)
\left(\begin{array}{cc}
  m_{44} & -m_{34} \\
  -m_{43} & m_{33} \\
\end{array}\right),
$$
where $\triangle_1=\det M_1$, $\triangle_2=\det M_2$. We observe
that
$$
b_{11} m_{44}-b_{12}m_{43}=0,
$$
$$
b_{22} m_{43}-b_{21}m_{44}=0,
$$
$$
b_{11} m_{12}-b_{21}m_{11}=0,
$$
$$
b_{22} m_{11}-b_{12}m_{12}=0.
$$
Taking the relations above into account we find
$$
(M_1^T)^{-1}BM_2^{-1}=\frac{\sqrt{2N(2N+\beta-1)}}{(c-1)\triangle_1\triangle_2}
\left(\begin{array}{cc}
  0 & -(m_{22}b_{11}-m_{21}b_{21})m_{34}+(m_{22}b_{12}-m_{21}b_{22})m_{33} \\
  0 & 0 \\
\end{array}\right).
$$
Now we will show that
$$
-(m_{22}b_{11}-m_{21}b_{21})m_{34}+(m_{22}b_{12}-m_{21}b_{22})m_{33}=\frac{\triangle_1\triangle_2}{\sqrt{c}}.
$$
Indeed, the straightforward algebra gives
\begin{equation}
\begin{split}
&-(m_{22}b_{11}-m_{21}b_{21})m_{34}+(m_{22}b_{12}-m_{21}b_{22})m_{33}\\
&=\sqrt{\beta+2N-1}\left[(\beta+2N)F(\beta,\beta+2N-1;\beta+2N;c)-2cNF(\beta,\beta+2N;\beta+2N+1;c)\right]\\
&\times\left[(\beta+2N-1)F(-2N,\beta-1;\beta;c)-2NF(-2N+1,\beta-1;\beta;c)\right]
\end{split}
\nonumber
\end{equation}
From the other hand,
$$
\triangle_1=\sqrt{\beta+2N-1}\left[(\beta+2N)F(\beta,\beta+2N-1;\beta+2N;c)-2cNF(\beta,\beta+2N;\beta+2N+1;c)\right],
$$
$$
\triangle_2=\sqrt{c}\left[(\beta+2N-1)F(-2N,\beta-1;\beta;c)-2NF(-2N+1,\beta-1;\beta;c)\right].
$$

Thus
$$
(M_1^T)^{-1}BM_2^{-1}=\frac{\sqrt{2N(2N+\beta-1)}}{(c-1)\sqrt{c}}\left(\begin{array}{cc}
                                                                         0 & 1 \\
                                                                         0 & 0
                                                                       \end{array}
\right).
$$
Formula (\ref{pf10}) follows after simple algebra.
\end{proof}
\textbf{Proof of Theorem \ref{MeixnerChSYMPLECTICTHEOREM}, a) and
Theorem
\ref{MeixnerChOOrtTheorem}, a)}.\\
 Formulae in the statements of the Theorem \ref{MeixnerChSYMPLECTICTHEOREM}, a) and Theorem
\ref{MeixnerChOOrtTheorem}, a) follow immediately from Corollary
\ref{COROLLARY1}, Corollary \ref{COROLLARY1ORTH}, and Proposition
\ref{PPPP}. \qed

\begin{prop}\label{PropDKCharlier}
In the case of the Charlier weight defined by equation
(\ref{ChW172}) we have
\begin{equation}
\left[D,K_N\right]=
\sqrt{\frac{2N}{a}}\left(\varphi_{2N-1}\otimes\varphi_{2N}+
\varphi_{2N}\otimes\varphi_{2N-1}\right).
\end{equation}
\end{prop}
\begin{proof}
According to Proposition (\ref{PropositionformulaDK}) the
commutation relation between the operators $D$ and $K_N$ is
determined by matrix elements of $2\times 2$ matrix $D_-$ defined
by equation (\ref{D+-}). This matrix can be found explicitly from
the system of difference equations for the orthonormal functions
$\{\varphi_n(x)\}_{n=0}^{\infty}$ associated with the Charlier
weight obtained in Proposition \ref{PSystemCh}. The result is
$$
D_-(x)=\left(\begin{array}{cc}
               \frac{x-2N}{a} & \frac{\sqrt{2N}}{a} \\
               -\sqrt{\frac{2N}{a}} & 1
             \end{array}
\right).
$$
Inserting the matrix elements of the matrix $D_-(x)$ into the
formula in the statement of Proposition
(\ref{PropositionformulaDK}), and taking into account that the
coefficient $a_{2N}$ in this formula equals $-\sqrt{2Na}$ we
obtain the desired result.
\end{proof}
\textbf{Proof of Theorem \ref{MeixnerChSYMPLECTICTHEOREM}, b) and
Theorem
\ref{MeixnerChOOrtTheorem}, b)}.\\
Since $\varphi_{2N-1}\in\mathcal{H}_N$, and
$\varphi_{2N}\in\mathcal{H}_N^{\perp}$ we can directly apply
Corollary \ref{COROLLARY1} and Corollary \ref{COROLLARY1ORTH}.
\qed
\section{A limiting relation between Meixner symplectic and orthogonal ensembles,
and the Charlier symplectic and orthogonal ensembles}
\begin{thm}\label{THEOREMLimitingRelations}
As $\beta\rightarrow\infty$ and $c=\frac{a}{\beta+a}$  the
correlation kernels for the Meixner symplectic and orthogonal
ensembles with weight $\frac{(\beta)_x}{x!}c^x$ (given in Theorem
\ref{MeixnerChSYMPLECTICTHEOREM}, a), Theorem
\ref{MeixnerChOOrtTheorem}, a) respectively) turn into the
correlation kernels for the Charlier symplectic and orthogonal
ensembles with weight $\frac{a^x}{x!}$ (given in in Theorem
\ref{MeixnerChSYMPLECTICTHEOREM}, b), Theorem
\ref{MeixnerChOOrtTheorem}, b) respectively).
\end{thm}
\begin{proof}
In order to prove  Theorem \ref{THEOREMLimitingRelations} we apply
the formula (\ref{MeixnerCHlimitrelation}). If
$\varphi^{Meixner}_n$ denotes the $\mbox{n\mbox{th}}$ orthonormal
function associated with the Meixner polynomials, and if
$\varphi^{Charlier}_n$ denotes the $\mbox{n\mbox{th}}$ orthonormal
function associated with the Charlier polynomials formula
(\ref{MeixnerCHlimitrelation}) implies
$$
\underset{\beta\rightarrow\infty}{\lim}\varphi_n^{Meixner}(x;\beta,\frac{a}{\beta+a})
=\varphi^{Charlier}_n(x;a).
$$
Therefore if $c=\frac{a}{\beta+a}$, and $\beta\rightarrow\infty$,
$$
K^{Meixner}_N(x,y)\simeq K^{Charlier}_N(x,y).
$$
Now the second statement of the Theorem can be checked
immediately, taking into account that the kernels of the operators
$\epsilon$ and $D$ in the case of the Meixner ensemble become
indistinguishable from the kernels of the corresponding operators
in the case of the Charlier ensemble, as $\beta\rightarrow\infty$
and $c=\frac{a}{\beta+a}$.
\end{proof}
\section{A limiting relation between the correlation functions of the Meixner and the Laguerre
symplectic ensembles} Fix the measure $\alpha_{Laguerre}$ on
$\R_{\geq 0}$ defined by
$$
\alpha_{Laguerre}(dx)=x^{\alpha}e^{-x}dx,\;\alpha>-1.
$$
Consider the set $\Conf_{N}(\R_{\geq 0})$ consisting of $N$-point
configurations $X=(x_1,\ldots, x_N)$. On this set we define a
probability measure $P_{N4}^{(\alpha)}$ as follows
$$
P_{N4}^{(\alpha)}(dX)=\const_{N4}|V(X)|^{4}\alpha_{Laguerre}^{\otimes}(dX),
$$
where
$\alpha_{Laguerre}^{\otimes}(dX)=\prod_{i=1}^N\alpha_{Laguerre}(dx_i)$,
$V(X)=\prod_{1\leq i<j\leq N}(x_i-x_j)$ is the Vandermonde
determinant,  and $\const_{N4}$ is the normalization constant.

If we interpret the points $x_1, x_2,\ldots, x_N$ of the random
point configuration $X$ as the eigenvalues of a $N\times N$
quaternion real  matrix then the measure $P_{N4}^{(\alpha)}$
determines the symplectic Laguerre ensemble of Random Matrix
Theory, see, for example, Mehta \cite{mehta}, Forrester
\cite{forrester0}.

Let $\{L_n^{(\alpha)}\}_{n=0}^{\infty}$ be the family of the
Laguerre  polynomials defined by the orthogonality relation
$$
\int\limits_0^{\infty}x^{\alpha}e^{-x}L_m^{(\alpha)}(x)L_n^{(\alpha)}(x)dx=\frac{\Gamma(\alpha+n+1)}{n!}\delta_{m,n},
$$
and set
\begin{equation}\label{16.0}
\varphi^{(\alpha)}_n(x)=\sqrt{\frac{n!}{\Gamma(\alpha+n+1)}}L_n^{(\alpha)}(x)x^{\frac{\alpha}{2}}e^{-\frac{x}{2}}.
\end{equation}
For the symplectic  Laguerre ensemble the correlation kernel is
the kernel of the operator $K_{N4}^{(\alpha)}$, which is of the
form (see, for example, Widom \cite{widom})
\begin{equation}\label{16.1}
K_{N4}^{(\alpha)}=\frac{1}{2}\left(\begin{array}{cc}
  \mathcal{D}S^{(\alpha)}_{N4} & \mathcal{D}S^{(\alpha)}_{N4}\mathcal{D} \\
  S^{(\alpha)}_{N4} & S^{(\alpha)}_{N4}\mathcal{D} \\
\end{array}\right),
\end{equation}
Here the operator $\Dc S^{(\alpha)}_{N4}$ has the kernel
\begin{equation}\label{DSalpha4}
\begin{split}
\Dc S^{(\alpha)}_{N4}(x,y)&=K_{N2}^{(\alpha)}(x,y)+\frac{\sqrt{2N(2N+\alpha)}}{2}(\sqrt{2N+\alpha}\zeta^{(\alpha)}_{2N}(x)-\sqrt{2N}\zeta^{(\alpha)}_{2N-1}(x))\\
&\times(\sqrt{2N}\Epsilon\zeta^{(\alpha)}_{2N}(y)-\sqrt{2N+\alpha}\Epsilon\zeta^{(\alpha)}_{2N-1}(y)),
\end{split}
\end{equation}
and the kernels of $S^{(\alpha)}_{N4}$,
$S^{(\alpha)}_{N4}\mathcal{D}$, and
$\mathcal{D}S^{(\alpha)}_{N4}\mathcal{D}$ can be obtained by
action of $\mathcal{D}$ and $\Epsilon$. In the formulae written
above the function $K^{(\alpha)}_{N2}(x,y)$ (which is the
correlation kernel for the unitary Laguerre ensemble of Random
Matrix Theory) is given by the formula
$$
K_{N2}^{(\alpha)}(x,y)=
-\sqrt{2N(2N+\alpha)}\frac{\varphi^{(\alpha)}_{2N}(x)\varphi^{(\alpha)}_{2N-1}(y)-
\varphi^{(\alpha)}_{2N-1}(x)\varphi^{(\alpha)}_{2N}(y)}{x-y},
$$
$\Dc$ is the operator of the differentiation, $\Epsilon$ is the
operator with the kernel $\Epsilon(x,y)=\frac{1}{2}\sgn(x-y)$, and
$\zeta_k^{(\alpha)}(x)=x^{-1}\varphi_k^{(\alpha)}(x)$. Note that our
notation is slightly different from that of Widom \cite{widom}.

It is clear from the definition of the symplectic Laguerre ensemble,
and from the definition of the symplectic Meixner ensemble that
there is a limiting relation between correlation functions of these
ensembles. Namely, denote by $\rho(x_1,\ldots,x_m)$ the correlation
function of the Meixner symplectic ensemble with weight
$w_{Meixner}(x)$ given by formula \ref{MeinerWeight171}, and by
$\rho^{(\alpha)}_m(x_1,\ldots,x_m)$ the correlation function of the
Laguerre symplectic ensemble defined by weight $x^{\alpha}e^{-x}$,
$x>0$. Set $\beta=1+\alpha$. Then
\begin{equation}\label{rhoLaguerrerhoMeixner}
\underset{c\rightarrow
1^{-}}{\lim}\;\frac{1}{(1-c)^m}\rho_m\left(\frac{x_1}{1-c},\ldots,\frac{x_m}{1-c}\right)
=\rho^{(\alpha)}_m\left(x_1,\ldots,x_m\right).
\end{equation}
Our aim here is to check equation (\ref{rhoLaguerrerhoMeixner}) on
the level of the correlation kernels.
\begin{thm}
Take $\beta=\alpha+1$. Then for any strictly positive integers
$x, y$ we have
$$
\underset{c\rightarrow
1^{-}}{\lim}\;\frac{1}{1-c}DS_{N4}\left(\frac{x}{1-c},
\frac{y}{1-c}\right) =\Dc S_{N4}^{(\alpha)}(x,y),
$$
$$
\underset{c\rightarrow 1^{-}}{\lim}\;S_{N4}\left(\frac{x}{1-c},
\frac{y}{1-c}\right) =\frac{1}{2}S_{N4}^{(\alpha)}(x,y),
$$
$$
\underset{c\rightarrow
1^{-}}{\lim}\;\frac{1}{1-c}(\nabla_+S_{N4})\left(\frac{x}{1-c},
\frac{y}{1-c}\right) =\frac{1}{2}\Dc S_{N4}^{(\alpha)}(x,y),
$$
$$
\underset{c\rightarrow
1^{-}}{\lim}\;-\frac{1}{1-c}(S_{N4}\nabla_-)\left(\frac{x}{1-c},
\frac{y}{1-c}\right) =\frac{1}{2}\Dc
S_{N4}^{(\alpha)}(y,x)=\frac{1}{2}S_{N4}^{(\alpha)}\Dc(x,y),
$$
$$
\underset{c\rightarrow
1^{-}}{\lim}\;-\frac{1}{(1-c)^2}(\nabla_+S_{N4}\nabla_-)\left(\frac{x}{1-c},
\frac{y}{1-c}\right) =\frac{1}{2}\Dc S_{N4}^{(\alpha)}\Dc(x,y).
$$
\end{thm}
\textit{Sketch of the proof.} The Meixner polynomials
$$
M_n(x;\beta,c)=F(-n,-x;\beta;1-{c}^{-1})
$$
are related with the Laguerre polynomials $L^{(\alpha)}_n(x)$ via
the formula
$$
\underset{c\rightarrow
1^{-}}{\lim}M_n\left(\frac{x}{1-c};\alpha+1,c\right)
=\frac{n!}{(\alpha+1)_n}L_n^{(\alpha)}(x).
$$

In addition, noting that
$$
\frac{(\beta)_{\frac{x}{1-c}}}{\Gamma(1+\frac{x}{1-c})}=\frac
{\Gamma(\beta+\frac{x}{1-c})}{\Gamma(\beta)\Gamma(1+\frac{x}{1-c})}
\simeq\left(\frac{x}{1-c}\right)^{\beta-1}\frac{1}{\Gamma(\beta)}\quad\mbox{as}\;
c\rightarrow 1,
$$
it is not hard to check that
\begin{equation}\label{ZVEZDA}
\underset{c\rightarrow
1^{-}}{\lim}\frac{1}{\sqrt{1-c}}\varphi_n\left(\frac{x}{1-c};
\alpha+1, c\right) =\varphi_n^{(\alpha)}(x),
\end{equation}
where $\varphi_n^{(\alpha)}(x)$ is defined by formula
(\ref{16.0}).  Now observe that
$\int\limits_0^{+\infty}\psi_1^{(\alpha)}(x)dx=0$, i.e.
\begin{equation*}
(\Epsilon\psi_1^{(\alpha)})(y)
=\frac{1}{2}\int\limits_{0}^y\psi_1^{(\alpha)}(x)dx-\frac{1}{2}\int\limits_y^{+\infty}\psi_1^{(\alpha)}(x)dx
=\int\limits_0^y\psi^{(\alpha)}_1(x)dx
=-\int\limits_{y}^{+\infty}\psi_1^{(\alpha)}(x)dx.
\end{equation*}
Indeed, for $\alpha>0$ $\Epsilon\psi_1^{(\alpha)}$ is an element
of $\Hc^{(\alpha)}_N$ ($\Hc^{(\alpha)}_N$ is spanned by
$\varphi^{(\alpha)}_0,\varphi_1^{(\alpha)},\ldots
,\varphi_{2N-1}^{(\alpha)}$), and we conclude that
$\Epsilon\psi_1^{(\alpha)}(0)=0$ for $\alpha>0$. But
$\Epsilon\psi_1^{(\alpha)}(0)=-\frac{1}{2}\int\limits_{0}^{+\infty}\psi_1^{(\alpha)}(x)dx$.
To check that $\int\limits_0^{+\infty}\psi_1^{(\alpha)}(x)dx=0$
holds true for $-1<\alpha\leq 0$ we can use the analytic
continuation with respect to the parameter $\alpha$. Using these
formulae we arrive to the  following
\begin{lem}\label{LemmaLimitingRelation}
For any two positive integers $x, y$
\begin{equation}
\underset{c\rightarrow 1^{-}}{\lim}\frac{1}{1-c}
K_N\left(\frac{x}{1-c},\frac{y}{1-c}\right)=K_{N2}^{(\alpha)}(x,y),
\nonumber
\end{equation}
$$
\underset{c\rightarrow
1^{-}}{\lim}\;\frac{1}{(1-c)^{3/2}}\psi_1\left(\frac{x}{1-c}\right)=
\sqrt{2N}\zeta_{2N}^{(\alpha)}(x)-\sqrt{2N+\alpha}\zeta^{(\alpha)}_{2N-1}(x)=\psi_1^{(\alpha)}(x),
$$
$$
\underset{c\rightarrow
1^{-}}{\lim}\frac{1}{(1-c)^{3/2}}\psi_2\left(\frac{x}{1-c}\right)=
\sqrt{2N+\alpha}\zeta^{(\alpha)}_{2N}(x)-\sqrt{2N}\zeta_{2N-1}^{(\alpha)}(x)=\psi_2^{(\alpha)}(x),
$$
$$
\underset{c\rightarrow
1^{-}}{\lim}\;\frac{1}{\sqrt{1-c}}\biggl(\epsilon\psi_1\biggr)\left(\frac{y}{1-c}\right)=\frac{1}{2}\left(\Epsilon\psi_1^{(\alpha)}\right)(y).
$$
\end{lem}
Using the limiting relations in the Lemma just stated above, and
the expression for $DS_{N4}$ in Theorem
\ref{MeixnerChSYMPLECTICTHEOREM}, a) we obtain
$$
\underset{c\rightarrow
1^{-}}{\lim}\;\frac{1}{1-c}DS_{N4}\left(\frac{x}{1-c},
\frac{y}{1-c}\right) =\Dc S_{N4}^{(\alpha)}(x,y).
$$
This is the first limiting relation between the kernels of the
Meixner symplectic ensemble and the Laguerre symplectic ensemble
stated in the Theorem. Other limiting relations can be obtained by
 action of operators $\nabla_{\pm}$ and $\epsilon$, and by application
 of Lemma \ref{LemmaLimitingRelation}. \qed

\section{Correlation functions for the Meixner orthogonal ensemble
and the parity respecting correlations for the Laguerre orthogonal
ensemble} Consider the Laguerre orthogonal ensemble. This ensemble
can be defined by the probability density function
$$
\const\cdot\prod\limits_{i=1}^{2N}e^{-\frac{z_i}{2}}z_i^{\frac{\alpha}{2}}\prod\limits_{1\leq
j<k\leq 2N}(z_j-z_k),
$$
where $0\leq z_1<z_2<\ldots <z_{2N}$. Denote by $x_1,\ldots ,x_N$
the odd labelled particles with respect to this ordering, and by
$y_1,\ldots, y_N$ the even labelled particles with respect to this
ordering.  We will also denote the probability density function
for this ensemble  by
$p^{(\alpha)}(x_1,\ldots,x_N;y_1,\ldots,y_N)$. Thus
$$
p^{(\alpha)}(x_1,\ldots,x_N;y_1,\ldots,y_N)=\const\cdot\prod\limits_{i=1}^{2N}e^{-\frac{z_i}{2}}z_i^{\frac{\alpha}{2}}\prod\limits_{1\leq
j<k\leq 2N}(z_j-z_k).
$$
The $(k_1,k_2)$-point correlation function for $k_1$ odd labelled
particles, and $k_2$ of even-labelled particles is defined as
\begin{multline*}
\rho_{(k_1,k_2)}^{(\alpha)}(x_1,\ldots,x_{k_1};y_1,\ldots,y_{k_2})=
\frac{N!}{(N-k_1)!}\,\frac{N!}{(N-k_2)!}\\\times\int_{(0,+\infty)^{k_1}}
\int_{(0,\infty)^{k_2}}p^{(\alpha)}(x_1,\ldots,x_N;y_1,\ldots,y_N)\prod\limits_{l=k_1+1}^N\prod\limits_{s=k_2+1}^N
dx_ldy_s.
\end{multline*}

As it is explained in Forrester and Rains \cite{forrester2},
$\rho_{(k_1,k_2)}^{(\alpha)}(x_1,\ldots,x_{k_1};y_1,\ldots,y_{k_2})$
describes parity respecting correlations of particles from the
Laguerre orthogonal ensemble, see \cite{forrester2} for details and
further references.
 Let $\rho_{(k_1,k_2)}(x_1,\ldots,
x_{k_1},y_1,\ldots, y_{k_2})$ be similarly defined correlation
function for the Meixner orthogonal ensemble. Let $c\to 1^-$ and
assume that
$$
\left[\frac{x_1}{1-c}\right],\ldots,
\left[\frac{x_{k_1}}{1-c}\right] \text{  are even,  }\quad
\left[\frac{y_1}{1-c}\right],\ldots,
\left[\frac{y_{k_2}}{1-c}\right] \text{  are odd}.
$$ Then definitions of  the Meixner and Laguerre orthogonal ensembles
imply
\begin{multline} \underset{c\rightarrow
1^{-}}{\lim}\;\frac{1}{(1-c)^{k_1+k_2}}\rho_{(k_1,k_2)}\left(\left[\frac{x_1}{1-c}\right],\ldots,
\left[\frac{x_{k_1}}{1-c}\right];
\left[\frac{y_1}{1-c}\right],\ldots,\left[\frac{y_{k_2}}{1-c}\right]
\right)\\
=\rho_{(k_1,k_2)}^{(\alpha)}(x_1,\ldots,x_{k_1};y_1,\ldots,y_{k_2}).
\end{multline}

\begin{thm}
\label{th:last} We have
\begin{multline}
\rho_{(k_1,k_2)}^{(\alpha)}(x_1,\ldots,x_{k_1};y_1,\ldots,y_{k_2})
\\= \Pf \left[\begin{array}{cc}
  \left[K_{N1}^{(\alpha)}(x_j,x_l)\right]_{j,l=1,\ldots,k_1}^{ee} & \left[K_{N1}^{(\alpha)}(x_j,y_l)\right]_{j=1,\ldots,k_1; l=1,\ldots,k_2}^{eo} \\
  \left[K_{N1}^{(\alpha)}(y_j,x_l)\right]_{j=1,\ldots,k_2;l=1,\ldots,k_1}^{oe} & \left[K_{N1}^{(\alpha)}(y_j,y_l)\right]_{j,l=1,\ldots,k_2}^{oo}\\
\end{array}\right],
\nonumber
\end{multline}
where
$$
\left[K_{N1}^{(\alpha)}(x,y)\right]^{ee}=\left[\begin{array}{cc}
  \left[\Epsilon S_{N1}^{(\alpha)}\right]^{ee}(x,y) & \left[S_{N1}^{(\alpha)}\right]^{ee}(x,y) \\
  \left[\Epsilon S_{N1}^{(\alpha)}\Epsilon-\Epsilon\right]^{ee}(x,y)
  & \left[S_{N1}^{(\alpha)}\Epsilon\right]^{ee}(x,y) \\
\end{array}\right],
$$
$$
\left[K_{N1}^{(\alpha)}(x,y)\right]^{oe}=\left[\begin{array}{cc}
  \left[\Epsilon S_{N1}^{(\alpha)}\right]^{oe}(x,y) & \left[S_{N1}^{(\alpha)}\right]^{oe}(x,y) \\
  \left[\Epsilon S_{N1}^{(\alpha)}\Epsilon-\Epsilon\right]^{oe}(x,y) &
   \left[S_{N1}^{(\alpha)}\Epsilon\right]^{oe}(x,y) \\
\end{array}\right],
$$
$$
\left[K_{N1}^{(\alpha)}(x,y)\right]^{eo}=\left[\begin{array}{cc}
  \left[\Epsilon S_{N1}^{(\alpha)}\right]^{eo}(x,y) &
  \left[S_{N1}^{(\alpha)}\right]^{eo}(x,y) \\
  \left[\Epsilon S_{N1}^{(\alpha)}\Epsilon-\Epsilon\right]^{eo}(x,y) &
  \left[S_{N1}^{(\alpha)}\Epsilon\right]^{eo}(x,y) \\
\end{array}\right],
$$
$$
\left[K_{N1}^{(\alpha)}(x,y)\right]^{oo}=\left[\begin{array}{cc}
  \left[\Epsilon S_{N1}^{(\alpha)}\right]^{oo}(x,y) &
  \left[S_{N1}^{(\alpha)}\right]^{oo}(x,y) \\
  \left[\Epsilon S_{N1}^{(\alpha)}\Epsilon-\Epsilon\right]^{oo}(x,y)
  & \left[S_{N1}^{(\alpha)}\Epsilon\right]^{oo}(x,y) \\
\end{array}\right],
$$
and
$$
\left[\Epsilon
S_{N1}^{(\alpha)}\right]^{ee}(x,y)=K_N^{(\alpha)}(x,y)+\sqrt{2N(2N+\alpha)}\left(\Epsilon^{e}
\psi_2^{(\alpha)}\right)(x)\psi_1^{(\alpha)}(y)=\left[\Epsilon
S_{N1}^{(\alpha)}\right]^{eo}(x,y),
$$
$$
\left[\Epsilon
S_{N1}^{(\alpha)}\right]^{oe}(x,y)=K_N^{(\alpha)}(x,y)+\sqrt{2N(2N+\alpha)}\left(\Epsilon^{o}
\psi_2^{(\alpha)}\right)(x)\psi_1^{(\alpha)}(y)=\left[\Epsilon
S_{N1}^{(\alpha)}\right]^{oo}(x,y),
$$
$$
\left[
S_{N1}^{(\alpha)}\Epsilon\right]^{ee}(x,y)=K_N^{(\alpha)}(x,y)+\sqrt{2N(2N+\alpha)}\left(\Epsilon^{e}
\psi_2^{(\alpha)}\right)(y)\psi_1^{(\alpha)}(x)=\left[
S_{N1}^{(\alpha)}\Epsilon\right]^{oe}(x,y),
$$
$$
\left[
S_{N1}^{(\alpha)}\Epsilon\right]^{eo}(x,y)=K_N^{(\alpha)}(x,y)+\sqrt{2N(2N+\alpha)}\left(\Epsilon^{o}
\psi_2^{(\alpha)}\right)(y)\psi_1^{(\alpha)}(x)=\left[
S_{N1}^{(\alpha)}\Epsilon\right]^{oo}(x,y),
$$
$$
\left[ \Epsilon
S_{N1}^{(\alpha)}\Epsilon-\Epsilon\right]^{ee}(x,y)=\left[\Epsilon^{e}K_N^{(\alpha)}\right](x,y)
-\Epsilon^{e}(x,y) +\sqrt{2N(2N+\alpha)}\left(\Epsilon^{e}
\psi_1^{(\alpha)}\right)(x)\left(\Epsilon^{e}
\psi_2^{(\alpha)}\right)(y),
$$
$$
\left[ \Epsilon
S_{N1}^{(\alpha)}\Epsilon-\Epsilon\right]^{eo}(x,y)=\left[\Epsilon^{e}K_N^{(\alpha)}\right](x,y)
-\Epsilon^{e}(x,y) +\sqrt{2N(2N+\alpha)}\left(\Epsilon^{e}
\psi_1^{(\alpha)}\right)(x)\left(\Epsilon^{o}
\psi_2^{(\alpha)}\right)(y),
$$
$$
\left[ \Epsilon
S_{N1}^{(\alpha)}\Epsilon-\Epsilon\right]^{oe}(x,y)=\left[\Epsilon^{o}K_N^{(\alpha)}\right](x,y)
-\Epsilon^{o}(x,y) +\sqrt{2N(2N+\alpha)}\left(\Epsilon^{o}
\psi_1^{(\alpha)}\right)(x)\left(\Epsilon^{e}
\psi_2^{(\alpha)}\right)(y),
$$
$$
\left[ \Epsilon
S_{N1}^{(\alpha)}\Epsilon-\Epsilon\right]^{oo}(x,y)=\left[\Epsilon^{o}K_N^{(\alpha)}\right](x,y)
-\Epsilon^{o}(x,y) +\sqrt{2N(2N+\alpha)}\left(\Epsilon^{o}
\psi_1^{(\alpha)}\right)(x)\left(\Epsilon^{o}
\psi_2^{(\alpha)}\right)(y),
$$
$$
\left[ S_{N1}^{(\alpha)}\right]^{ee}(x,y)=\frac{\partial}{\partial
x}K_N^{(\alpha)}(x,y)+\frac{\sqrt{2N(2N+\alpha)}}{2}
\psi_2^{(\alpha)}(x)\psi_1^{(\alpha)}(y),
$$
$$
\left[ S_{N1}^{(\alpha)}\right]^{ee}(x,y)=\left[
S_{N1}^{(\alpha)}\right]^{oe}(x,y)=\left[
S_{N1}^{(\alpha)}\right]^{eo}(x,y)=\left[
S_{N1}^{(\alpha)}\right]^{oo}(x,y).
$$
In the formulae above the operators $\Epsilon^{e}$ and
$\Epsilon^{o}$ are defined by the relations
$$
\left(\Epsilon^{e}f\right)(x)=-\frac{1}{2}\int_{x}^{+\infty}f(y)dy,\;\;\left(\Epsilon^{o}f\right)(x)=\frac{1}{2}\int^{x}_{0}f(y)dy.
$$
\end{thm}
\begin{proof}
The correlation kernel $K_{N1}$ for the Meixner orthogonal ensemble
is obtained explicitly in Theorem \ref{MeixnerChOOrtTheorem}, a). To
find the kernel for the correlation function
$\rho_{(k_1,k_2)}^{(\alpha)}$ we need to compute
$K_{N1}([\frac{x}{1-c}],[\frac{y}{1-c}])$ as $c\rightarrow 1^{-}$.
This can be done exploiting formulae
 in Lemma \ref{LemmaLimitingRelation}, and in addition the
 following limiting relations
 $$
\underset{c\rightarrow
1^{-}}{\lim}\;\frac{1}{\sqrt{1-c}}\left(\epsilon\psi_2\right)
\left(\left[\frac x{1-c}\right]\right)\
=\left\{%
\begin{array}{ll}
    \left(\Epsilon^{e}\psi_2^{(\alpha)}\right)(x), & \left[\frac x{1-c}\right]\;\hbox{is even,} \\
    \left(\Epsilon^{o}\psi_2^{(\alpha)}\right)(x), & \left[\frac x{1-c}\right]\; \hbox{is odd,} \\
\end{array}%
\right.
$$
$$
\underset{c\rightarrow
1^{-}}{\lim}\;\left(\epsilon K_{N}\right)
\left(\left[\frac{x}{1-c}\right],\left[\frac{y}{1-c}\right]\right)=\left\{%
\begin{array}{ll}
    \left(\Epsilon^{e}K_N^{(\alpha)}\right)(x,y), & \left[\frac x{1-c}\right]\;
    \hbox{is even,} \\
    \left(\Epsilon^{o}K_N^{(\alpha)}\right)(x,y), & \left[\frac x{1-c}\right]\;
    \hbox{is odd.} \\
\end{array}%
\right.
$$
\end{proof}

Observe that the structure of the parity respecting correlation
functions in Theorem \ref{th:last} is very similar to that in
Section 3.1 of \cite{forrester2}. For $\alpha=0$ the kernels of
Theorem \ref{th:last} and the case $A=0$ of \cite{forrester2} must
be equivalent. However, direct verification of this fact is not an
easy task, see e.g. Section 5.1 of \cite{forrester2}, and we
postpone the discussion of this equivalence until a later
publication.

\end{document}